\documentclass[lettersize,journal]{IEEEtran}

\usepackage[utf8]{inputenc}
\usepackage[T1]{fontenc}
\usepackage{amsmath,amsfonts,amssymb}
\usepackage{algorithmic}
\usepackage{algorithm}
\usepackage{array}
\usepackage{textcomp}
\usepackage{stfloats,balance}
\usepackage{url,comment}
\usepackage{graphicx}
\usepackage{tikz}
\usetikzlibrary{arrows.meta,positioning,fit,backgrounds,calc}
\usepackage{cite,xcolor}
\graphicspath{{figures/}}
\DeclareMathOperator*{\argmax}{arg\,max}
\usepackage{fancyhdr}

\fancypagestyle{preprintnotice}{%
  \fancyhf{}%
  \fancyhead[C]{%
    \footnotesize
    This work has been submitted to the IEEE for possible publication.
    Copyright may be transferred without notice, after which this version
    may no longer be accessible.%
  }%
}

\begin{document}

\title{Twin-Fidelity-Aware Resolution of Direct\\ xApp Conflicts in Open RAN}

\author{Akram~Almohammedi,~\IEEEmembership{Member,~IEEE}, 
Mohammed~Balfaqih,~\IEEEmembership{Senior Member,~IEEE}, Sam~Darshi,~\IEEEmembership{Senior Member,~IEEE},
Rami~Langar,~\IEEEmembership{Member,~IEEE}, and Wael~Jaafar,~\IEEEmembership{Senior Member,~IEEE}%
\thanks{A.~Almohammedi is with the Department of Software and IT Engineering, \'Ecole de Technologie Sup\'erieure (\'ETS), University of Quebec, Montreal, Canada (e-mail: akram.al-zaghir@etsmtl.ca).}%
\thanks{Mohammed Balfaqih is with the Department of Computer and Network Engineering, University of Jeddah, Jeddah 23890, Saudi Arabia (e-mail: mabalfaqih@uj.edu.sa).}%
\thanks{Sam Darshi is with the Department of Electrical Engineering, Indian Institute of Technology Ropar, Ropar 140001, India (e-mail: sam@iitrpr.ac.in).}%
\thanks{R.~Langar and W.~Jaafar are with the Department of Software and IT Engineering, \'Ecole de Technologie Sup\'erieure (\'ETS), University of Quebec, Montreal, Canada (e-mail: \{rami.langar, wael.jaafar\}@etsmtl.ca). R.~Langar is also with LIGM-CNRS UMR~8049, University Gustave Eiffel, F-77420 Marne-la-Vall\'ee, France. \\ This work is funded by the FRQ Science in Exile program (FRQNT 2026-EXICX-380348).}}

\maketitle
\thispagestyle{preprintnotice}

\begin{abstract}
Open Radio Access Network (O-RAN) allows independently developed xApps to control RAN functions through the Near-Real-Time RAN Intelligent Controller (Near-RT RIC). When multiple xApps with conflicting objectives operate concurrently, they may issue incompatible control actions that degrade network performance. This paper addresses a direct conflict in which an energy-saving (ES) xApp and a coverage/throughput-oriented (CTO) xApp request different downlink transmit-power settings for the same cell. We formulate conflict resolution as the online selection of a continuous blend of the two proposals, maximizing an energy-aware utility that jointly considers throughput and power consumption. A network digital twin (NDT) predicts this utility for candidate actions before live deployment. However, relying solely on the highest twin-predicted utility becomes ineffective when the twin drifts. To address this issue, we propose a twin-fidelity-aware hard-switching arbiter that continuously monitors the prediction error between predicted and observed utilities using an exponentially weighted moving average. When the prediction error remains below a threshold, the arbiter follows the NDT-selected action; otherwise, it switches to the best previously observed action learned online. The proposed arbiter is lightweight, training-free, and requires no oracle knowledge of the optimal policy. System-level 5G evaluations show that it consistently achieves the closest throughput-power trade-off to the optimum across different operator energy priorities, yielding the lowest normalized utility regret ($0.017 \pm 0.006$, versus $0.159 \pm 0.052$ for a COMIX-style twin-based selector). Under severe NDT drift (10~dB), it further reduces the utility regret from $11.19 \pm 3.58$ to $0.55 \pm 0.25$. These results demonstrate that explicit online monitoring of twin fidelity enables robust digital-twin-assisted xApp conflict resolution while preserving utility-aware throughput-power optimization.
\end{abstract}


\section{Introduction}\label{sec:intro}
\IEEEPARstart{T}{he} Open Radio Access Network (O-RAN) disaggregates the radio access network into interoperable, software-driven components connected through open, standardized interfaces \cite{ref1}. This openness enhances interoperability, promotes vendor diversity, and enables network operators to combine components from different vendors without sacrificing functionality. Central to this architecture is the RAN Intelligent Controller (RIC), which opens network control to independently developed applications. In the Near-Real-Time RIC (Near-RT RIC), these applications, called xApps, monitor network state and issue control actions on control timescales between 10~ms and 1~s \cite{ref1}. This programmability enables flexible, multi-vendor RAN optimization, but it also introduces a new control problem. Indeed, xApps developed in isolation, often by different vendors, may pursue conflicting objectives and issue control actions that interfere with one another \cite{ref2}.

Conflicts among xApps are commonly categorized as direct, indirect, and implicit \cite{ref2,ref3}. A direct conflict occurs when two or more xApps attempt to set the same control parameter to different values. An indirect conflict arises when distinct parameters controlled by different xApps affect a common key performance indicator (KPI), while an implicit conflict emerges through less obvious dependencies that may become evident only during operation \cite{ref2,ref3}. Among these, the direct conflict is the most explicit form of contention and poses the most immediate arbitration problem. In fact, before any action reaches the RAN, the Near-RT RIC must decide which value or what compromise to apply to the shared parameter.

Within the direct conflict context, this paper addresses a particular conflict between an energy-saving (ES) xApp and a coverage/throughput-oriented (CTO) xApp that both control a cell's downlink transmit power, a setting motivated by prior O-RAN conflict studies in which energy-saving and mobility-, throughput-, or coverage-oriented xApps issue opposing transmit-power decisions \cite{ref4,ref5}. The ES xApp lowers transmit power to reduce energy consumption and inter-cell interference, whereas the CTO xApp raises it to improve received signal quality, coverage, and aggregate throughput. The two proposals cannot be reconciled using a static policy, as applying either one alone sacrifices the competing objective, while a fixed compromise fails to adapt to varying network conditions and operator energy-throughput preferences. Hence, the arbitration problem is to select, at runtime, a continuous blend weight between the ES and CTO proposals, where the blend determines the applied downlink transmit power according to both the current network state and the operator's energy-throughput priority.

An effective manner to arbitrate such conflicts is to evaluate candidate actions on a network digital twin (NDT) before applying them to the live network. NDTs are regarded as enablers of intelligent and resilient O-RAN, supporting network modeling, artificial intelligence and machine learning (AI/ML) training and testing, performance assurance, network planning, and energy-saving \cite{ref6,ref7}. COMIX \cite{ref5} follows this direction, using an NDT to evaluate conflicting xApp actions before committing a final decision to the live network. However, such an approach is only as reliable as the twin itself. In practice, a twin may drift from the live network through modeling error, stale calibration, and changing interference and traffic conditions. Moreover, O-RAN digital-twin studies stressed that model-based solutions require validation and performance assurance, as their behavior varies across environments and may harm the live network if left untested \cite{ref7}. A twin-based arbiter that does not account for this possibility silently inherits the twin's errors precisely when accuracy matters most.

Prior O-RAN conflict-management research spans conflict detection and characterization~\cite{ref2,ref9,ref12}, mitigation through prioritization or QoS-aware optimization~\cite{ref4,ref8}, learning-based coordination or consolidation~\cite{ref16,ref17}, and twin-assisted action selection~\cite{ref5}. A complementary research direction investigates digital twins and runtime assurance for RAN control~\cite{ref7,ref22,ref23}; Section~II reviews both directions.
Within this body of work, existing mitigation methods either presume that the digital twin or learned model remains reliable throughout operation (e.g.,~\cite{ref5,ref21}), optimize a different objective such as quality-of-service (QoS)-threshold satisfaction (e.g.,~\cite{ref4,ref8}), or require offline training, coordination, or modification of the deployed xApps (e.g.,~\cite{ref16,ref17}). However, a lightweight, training-free runtime arbiter that explicitly accounts for NDT drift during direct conflict resolution is still lacking.


Motivated by the above, we propose a twin-fidelity-aware arbitration method for direct xApp conflict resolution. Indeed, rather than treating the NDT as always trustworthy, the proposed arbiter continuously estimates whether the twin's predictions remain consistent with live network observations or not. When the measured fidelity is high, the arbiter exploits the twin-selected action. However, if fidelity degrades, the arbiter falls back to a last-known-good action proven on the live network during operation. Hence, the Near-RT RIC retains the predictive benefit of the NDT without depending blindly on its estimates that may no longer reflect the reality of the network. Consequently, the main contributions of this paper are as follows:
\begin{itemize}
\item We formulate direct ES/CTO transmit-power conflict resolution as the online selection of a continuous blend weight between the two xApp proposals. The selected blend sets the cell's downlink transmit power and is evaluated through an energy-aware utility that captures the throughput-power trade-off.
\item We propose a training-free twin-fidelity-aware hard-switching arbiter. The arbiter maintains an exponentially weighted moving average (EWMA) of the mismatch between twin-predicted and live-observed utility. When this mismatch remains below a predefined threshold, the twin-selected action is applied; otherwise, the arbiter switches to an online-learned last-known-good action. Our method requires neither offline training nor oracle knowledge of the true optimum.
\item Through extensive experiments, we evaluate the proposed arbiter in a multi-cell system-level simulation and compare it to several baselines 
in terms of two complementary metrics, namely utility regret and QoS satisfaction. Across operator energy-priority settings and under NDT drift, explicit fidelity monitoring is shown to enhance robustness over blind twin-based selection, especially under NDT drift conditions. We further characterize our method's operating assumptions, namely a calibration warm-up and an approximately stationary live optimum.
\end{itemize}


The remainder of the paper is organized as follows. Section~\ref{sec:related} reviews the related work on conflict detection, mitigation and arbitration, digital-twin-based assurance, and practical deployment in O-RAN. Section~\ref{sec:model} presents the system model and Section~\ref{sec:problem} the problem formulation. Section \ref{sec:algo} details the proposed twin-fidelity-aware arbiter. Section~\ref{sec:eval} describes the evaluation methodology. Then, Section~\ref{sec:results} presents the results and discussion. Finally, Section~\ref{sec:conclusion} concludes the paper.

\section{Related Work}\label{sec:related}
Research on O-RAN xApp conflicts can be viewed along the conflict-management lifecycle, which motivates the four directions reviewed in this section. A conflict must first be \emph{detected and characterized} (Subsection~\ref{subsec:rw-detect}), then \emph{mitigated or arbitrated} by selecting, prioritizing, coordinating, or combining competing xApp actions (Subsection~\ref{subsec:rw-mitigate}). Because recent arbitration methods increasingly relied on NDTs or learned models, \emph{twin-based evaluation and runtime assurance} form a third direction (Subsection~\ref{subsec:rw-twin}). Finally, \emph{practical deployment} studies define the operational constraints a deployable solution must satisfy, including modularity, low overhead, and compatibility with independently developed xApps (Subsection~\ref{subsec:rw-practical}). The proposed method lies at the intersection of the second and third directions.
The first and fourth directions are not direct competitors but establish, respectively, the conflict model we adopt and the practical requirements that shape our design.

\subsection{Conflict Detection and Characterization}\label{subsec:rw-detect}
Programmable O-RAN enables independently developed xApps and rApps to act on shared or coupled RAN control parameters, creating the possibility of conflicting objectives. Following the O-RAN taxonomy, such conflicts are commonly categorized as direct, indirect, or implicit \cite{ref3}. Several works focused on detecting, profiling, and characterizing these conflicts before they degrade service. For instance, PACIFISTA \cite{ref9} is a sandbox-based conflict evaluation and management framework that profiled O-RAN applications under controlled operational conditions, detecting both direct conflicts and parameter- Key Performance Metric (KPM)-level conflicts, and quantifying their severity using statistical distance measures. It also provided threshold-based mitigation by deciding which applications should be allowed to coexist or be blocked before deployment. However, it did not perform fine-grained runtime arbitration among conflicting xApp control actions. 
Similarly, authors in \cite{ref10} identified policy conflicts among xApps as a major operational challenge in O-RAN environments. Their analysis examined direct, indirect, and implicit conflicts and highlighted their potential impact on network performance, stability, and security. The work further discussed the role of AI/ML techniques in supporting conflict detection and mitigation. However, its primary contribution lies in conflict taxonomy, risk analysis, and architectural considerations.
More recent studies have leveraged graph analytics, ML, and causal inference to capture complex dependencies among xApps, control parameters, and network KPIs. In \cite{ref11}, GraphSAGE was employed to reconstruct and classify O-RAN conflict graphs, enabling the discovery of latent relationships among applications and the resources they influence. GRAPHICA \cite{ref12} extended this idea through graph convolutional learning, predicting direct, indirect, and implicit conflicts while simultaneously identifying root-cause xApps, even under highly imbalanced conflict datasets. Also, the authors of \cite{ref13} combined explainable ML with causal-inference techniques to identify potentially conflicting RAN control parameters and estimate their causal impact on performance indicators, thus improving conflict interpretability and root-cause analysis. Finally, the architecture proposed in \cite{ref15} introduces a proactive conflict-screening mechanism within the Service Management and Orchestration (SMO)/Non-RT RIC that evaluates proposed rApp policy or configuration changes before they are introduced into the operational network. By identifying potential incompatibilities at design, such approaches reduce the likelihood of harmful interactions reaching the Near-RT RIC.



\subsection{Conflict Mitigation and Arbitration}\label{subsec:rw-mitigate}

Once conflicts have been identified, they must be resolved in a manner that preserves network performance while respecting the objectives of competing xApps. Existing conflict-mitigation approaches differ in how they arbitrate among conflicting control intents, ranging from rule-based prioritization and QoS-aware optimization to learning-based coordination and NDT-assisted decision making. 
In \cite{ref2}, the authors proposed the Conflict Mitigation Framework (CMF) embedded within the Near-RT RIC. It introduced dedicated conflict-detection and conflict-resolution components that monitor RAN control messages and identify direct, indirect, and implicit conflicts among xApps. The framework demonstrated conflict resolution through prioritization-based policies, whereby the control action of a higher-priority xApp supersedes competing actions. CMF established the feasibility of integrating conflict-management functionality directly within the Near-RT RIC and remains a key reference architecture for O-RAN conflict mitigation. 
Several studies formulated conflict mitigation as an optimization problem. For example, QoS-Aware Conflict Mitigation (QACM) \cite{ref8} introduced a QoS-aware conflict-management mechanism that selects shared control-parameter values to maximize the number of xApps whose QoS requirements are simultaneously satisfied. Similarly, \cite{ref4} investigated threshold-aware mitigation strategies for a direct conflict between energy-saving (ES) and mobility-robustness optimization (MRO) xApps, evaluating trade-offs among competing performance objectives under different QoS constraints. Building on this, the AI-powered framework proposed in \cite{ref14} leveraged Generation of Conflicts (GenC) to generate labeled conflict datasets and trained GNN-, Bi-LSTM-, and SMOTE-GNN-based models for scalable conflict detection and classification, integrating mitigation through a Conflict Management System (CMS)/QACM-oriented workflow. These approaches are particularly relevant because they consider competing objectives acting on shared control parameters. 
Conflicts have also been addressed through cooperation among learning agents or by consolidating control functionality. In \cite{ref16}, a team-learning framework enabled resource-allocation and power-allocation xApps to exchange intended actions and jointly optimize resource allocation, yielding improvements in throughput and packet-drop performance compared with independently operating agents. A different strategy is proposed in \cite{ref17}, where knowledge from multiple xApps is distilled into a single unified controller capable of reproducing their collective functionality while eliminating inter-xApp conflicts. 
Moreover, a context-aware scheduler was proposed in \cite{ref18}, where it dynamically determines which xApps should be activated on the basis of network context and target performance objectives, thus reducing conflicting interactions without requiring joint training of the participating applications. Similarly, \cite{ref19} investigated conflict mitigation under varying levels of information sharing and inter-vendor coordination, emphasizing the fundamental trade-off between mitigation effectiveness and the operational independence of third-party xApps. 
Finally, among existing approaches, COMIX \cite{ref5} is the closest to our work. COMIX resolved power-control conflicts by evaluating candidate xApp actions using an NDT and selecting the action predicted to yield the highest policy-specific performance. Although we leverage an NDT like COMIX, the latter assumes that the twin remains accurate for decision making throughout operation. In contrast, we explicitly model twin reliability as a time-varying runtime property and adapt the arbitration process to account for prediction errors and model drift.

\subsection{Digital Twins and Runtime Assurance}\label{subsec:rw-twin}
While the conflict-mitigation approaches discussed in Section~\ref{subsec:rw-mitigate} focus on determining \emph{which} action should be applied when xApps disagree, a complementary research direction investigates the \emph{trustworthiness} of the models and NDTs increasingly used to support such decisions. On the one hand, as O-RAN evolves toward AI-native operation, NDTs are emerging as key enablers for evaluating control actions, validating policies, and assessing network behavior before changes are applied to the live RAN. 
The NDT-enabled O-RAN vision proposed in \cite{ref6} advocates the use of digital twins for AI/ML training, network monitoring, policy optimization, and resilient closed-loop control in intelligent 6G RANs. Similarly, the O-RAN nGRG DT-RAN report \cite{ref7} identified a broad range of applications for digital twins, including AI/ML model development, performance assurance, network-testing automation, network planning, energy-efficiency optimization, and site-specific configuration tuning. Beyond conceptual frameworks, digital twin-assisted control loops have been explored for energy-aware resource management in O-RAN-based fixed wireless access networks \cite{ref20}, while \cite{ref21} integrated a digital twin with multi-agent reinforcement learning (MARL) to support intelligent xApp management in Internet-of-Things (IoT)-enabled O-RAN. These works establish the NDT as a valuable environment for experimentation, training, validation, and decision support.

On the other hand, as increasingly autonomous control functions are introduced into the RAN, model inaccuracies, distribution shifts, and unforeseen operating conditions can compromise decision quality. To address this issue, the Safety Copilot framework \cite{ref22} intercepts AI-generated Near-RT RIC control actions and verifies them against predefined safety constraints before they are enforced in the operational network. Similarly, AIDITA \cite{ref23} presented an end-to-end (E2E) AI-driven NDT platform for traffic analytics in O-RAN. By combining real-time KPM collection, GenAI-based synthetic-data generation, and incremental model updates, AIDITA continuously adapts traffic-prediction and anomaly-detection models under evolving network conditions. Although its primary objective is adaptive traffic analytics rather than xApp conflict management, it demonstrated the feasibility of continuously aligning an NDT environment with live-network observations.

\subsection{Practical Implementation and Industry Perspectives}\label{subsec:rw-practical}
Beyond algorithmic performance, several studies emphasize that conflict-management solutions must remain compatible with the open, multi-vendor nature of O-RAN. In practice, xApps and rApps are often developed by different vendors, operate under limited information sharing, and evolve independently over time. As a result, conflict-resolution mechanisms must not only achieve effective arbitration but also preserve modularity, interoperability, and ease of deployment. Experimental demonstrations have highlighted the operational importance of coordinating independently developed applications. The multi-vendor O-RAN RIC demonstration reported in \cite{ref24} integrated an energy-saving rApp with a traffic-steering xApp and showed that energy-efficiency objectives must be carefully coordinated with traffic-management decisions to avoid degrading user experience. Similarly, the study in \cite{ref25} analyzed practical lessons learned from the design, implementation, and evaluation of O-RAN xApps, identifying challenges related to platform integration, service-model availability, portability, testing, lifecycle management, and third-party application deployment. These studies suggest that deployable conflict-management solutions should operate as external coordination layers rather than requiring modifications to existing xApps or rApps. Also, industry-driven efforts further reinforce the significance of conflict management in operational O-RAN deployments. For instance, the \emph{RIC-Apps Conflict Management} white paper \cite{ref26} provided a comprehensive discussion of vertical and horizontal conflicts, direct, indirect, and implicit conflict classes, and the corresponding requirements from operators, platform providers, and application developers. It outlined practical conflict-detection mechanisms based on shared control parameters and KPI interactions, and highlighted the need for scalable conflict management in real-world RIC deployments. In addition, direct xApp conflict detection and mitigation have been experimentally demonstrated on a software-defined radio (SDR) O-RAN platform \cite{ref27} integrating srsRAN, Open5GS, USRP hardware, and an O-RAN Software Community Near-RT RIC. The results confirmed the practical feasibility of runtime conflict handling while illustrating the limitations of simple resolution policies when competing xApps reflect legitimate operator trade-offs, such as balancing energy efficiency against throughput or user-experience objectives.

As discussed in this section, existing solutions use techniques ranging from sandbox-based profiling, graph learning, and causal analysis to QoS-aware optimization, cooperative learning, scheduling, testbed validation, and NDT-assisted decision support. However, a gap remains between twin-assisted arbitration and deployable runtime operation. In particular, existing approaches do not provide a lightweight, training-free arbitration mechanism for direct conflicts over downlink transmit-power control when the reliability of the NDT itself changes over time. In this paper, we address this gap by continuously monitoring twin fidelity and adapting arbitration decisions accordingly, falling back to a live-validated action whenever the discrepancy between predicted and observed utility indicates that the twin is no longer sufficiently trustworthy.

To clearly contextualize our contributions, Table~\ref{tab:comparison}
summarizes these distinctions across representative conflict-management and
digital-twin-assisted O-RAN works, highlighting that the proposed arbiter
uniquely combines runtime NDT-in-the-loop decision making, explicit
NDT-fidelity monitoring, live-validated fallback, and deployment without
offline training or per-application profiling.
\begin{table*}[!t]
\caption{Comparison With Related O-RAN Conflict-Management and Digital-Twin Works}
\label{tab:comparison}
\centering
\renewcommand{\arraystretch}{1.2}
\footnotesize
\begin{tabular}{@{}p{3.0cm} p{4.1cm} c c c c c@{}}
\hline
\textbf{Work} & \textbf{Main focus} &
\begin{tabular}[c]{@{}c@{}}\textbf{Direct runtime}\\\textbf{arbitration}\end{tabular} &
\begin{tabular}[c]{@{}c@{}}\textbf{NDT in the}\\\textbf{decision loop}\end{tabular} &
\begin{tabular}[c]{@{}c@{}}\textbf{Runtime fidelity}\\\textbf{monitoring}\end{tabular} &
\begin{tabular}[c]{@{}c@{}}\textbf{No offline train-}\\\textbf{ing/profiling}\end{tabular} &
\begin{tabular}[c]{@{}c@{}}\textbf{Live-validated}\\\textbf{fallback}\end{tabular} \\
\hline
CMF~\cite{ref2} & Rule-based detection, priority resolution & Yes & No & No & Yes & No \\
ES/MRO mitigation~\cite{ref4} & Threshold/QoS mitigation of ES--MRO power conflict & Yes & No & No & Partial$^{\dagger}$ & No \\
QACM~\cite{ref8} & QoS-aware shared-parameter optimization & Yes & No & No & Partial$^{\dagger}$ & No \\
PACIFISTA~\cite{ref9} & Sandbox profiling, pre-deployment screening & Pre-deployment & No & No & No & No \\
GNN/causal detection~\cite{ref11,ref12,ref13} & Conflict detection, prediction, root cause & No & No & No & No & No \\
GenC/CMS~\cite{ref14} & Learned detection/classification + mitigation workflow & Partial & No & No & No & No \\
Team learning / distillation~\cite{ref16,ref17} & Joint training or xApp consolidation & Partial & No & No & No & No \\
Scheduling/coordination~\cite{ref18,ref19} & xApp activation, coordination levels & Partial & No & No & Partial$^{\ddagger}$ & No \\
COMIX~\cite{ref5} & NDT-assisted conflict-resolution policies & Yes & Yes & No & Yes & No \\
DT frameworks and runtime assurance~\cite{ref6,ref7,ref20,ref21,ref22,ref23}$^{\S}$ & Twin-enabled control, training, model alignment & No & Yes & Partial & No & No \\
\textbf{This work} & Twin-fidelity-aware ES/CTO power arbitration & \textbf{Yes} & \textbf{Yes} & \textbf{Yes} & \textbf{Yes} & \textbf{Yes} \\
\hline
\end{tabular}

\vspace{3pt}
\begin{minipage}{\textwidth}\footnotesize
$^{\dagger}$The QACM decision rule is optimization-based but presupposes
per-xApp KPI prediction models, instantiated as offline-trained regressors
in~\cite{ref8}; the same presupposition applies where~\cite{ref4}
deploys QACM as its mitigation method.
$^{\ddagger}$The scheduler in~\cite{ref18} is itself a trained agent,
although no joint training or retraining of the managed xApps is required.
$^{\S}$The runtime safeguards in~\cite{ref22} verify actions against
predefined safety constraints and may roll back to a previously verified
safe configuration or invoke a conservative rule-based controller, while
\cite{ref23} continuously realigns the twin with live observations.
Neither approach gates arbitration on an explicit twin-fidelity signal nor
provides a fidelity-triggered, online-learned, live-validated fallback
action. Likewise, reactive rollback to a previous or default configuration
(e.g.,~\cite{ref3,ref4}) is neither fidelity-triggered nor online-learned.
\end{minipage}
\end{table*}

\section{System Model}\label{sec:model}
We consider an O-RAN Near-RT RIC environment in which multiple
independently developed xApps share control over RAN parameters.
Such an architecture enables flexible network optimization but may also create conflicting control decisions when different xApps attempt to modify the same parameter. This work focuses on a direct conflict between an energy-saving (ES) xApp and a coverage/throughput-oriented (CTO) xApp that simultaneously control
the downlink transmit power of a cell. This section defines the
conflict scenario, the action space, and the utility metric used to evaluate candidate conflict-resolution decisions.

\subsection{Direct Conflict over a Shared Continuous Parameter}
\begin{figure*}[!t]
\centering
\includegraphics[width=0.8\linewidth]{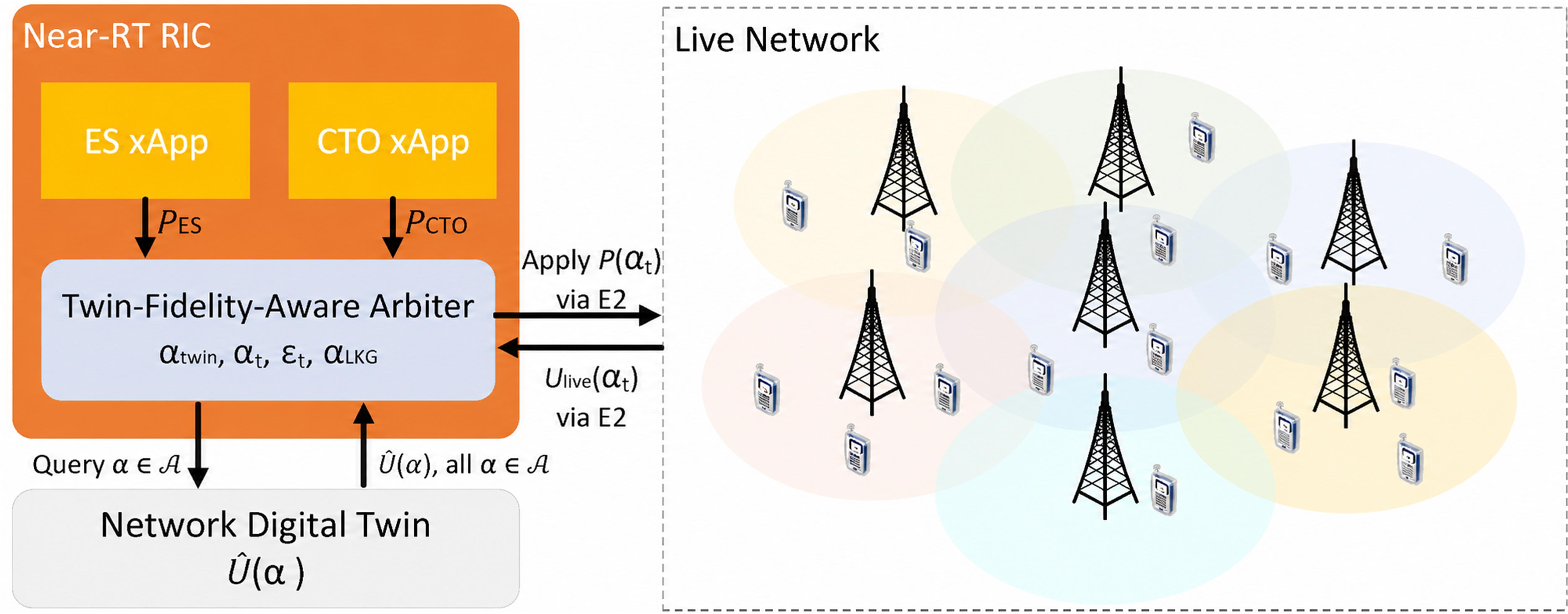}
\caption{System architecture of the twin-fidelity-aware hard-switching arbiter.}
\label{fig:arch}
\end{figure*}

Consider a Near-RT RIC hosting a set of xApps, denoted by $\mathcal{X}$. A direct conflict arises when multiple xApps attempt to control the same RAN parameter and issue incompatible control actions. In this paper, we consider the canonical two-xApp conflict in which both ES and CTO xApps simultaneously control the downlink transmit power of a cell of interest. The ES xApp favors lower transmit power to reduce energy consumption, whereas the CTO xApp favors higher transmit power to improve coverage and throughput. Executing either request directly prioritizes one objective at the expense of the other. Consequently, a conflict-arbitration mechanism is required to determine a single transmit-power value to be applied by the Near-RT RIC at each control interval. The system architecture is illustrated in Fig.~\ref{fig:arch}.


\subsection{Action Parameterization}

Let $P_{\mathrm{ES}}$ and $P_{\mathrm{CTO}}$ denote the transmit-power requests issued by the ES and CTO xApps, respectively, with $P_{\mathrm{ES}} < P_{\mathrm{CTO}}$.  The specific values used in the evaluation are reported in Subsection~\ref{subsec:scenario} and Table~\ref{tab:params}, although the formulation applies to any pair of conflicting power requests. The arbiter selects a blending factor $\alpha \in [0,1]$ and applies the transmit-power value, as follows:
\begin{equation}
\label{eq:1}
P(\alpha)
=
P_{\mathrm{ES}}
+
\alpha
\left(
P_{\mathrm{CTO}}
-
P_{\mathrm{ES}}
\right),
\end{equation}
where $P_{\mathrm{ES}}$, $P_{\mathrm{CTO}}$, and $P(\alpha)$ are expressed in dBm; hence, the interpolation is performed on the logarithmic power scale. Here, $\alpha=0$ reproduces the ES request and $\alpha=1$ reproduces the CTO request. Thus, $\alpha$ parameterizes a continuum of compromise actions between the two extremes. To simplify runtime arbitration, the continuous action space is discretized into the finite candidate set
\begin{equation}
\label{eq:2}
\mathcal{A}
=\{0,\,0.05,\,0.10,\,\ldots,\,1\}.
\end{equation}
Since the energy cost is naturally measured in watts (W), the selected transmit power is converted from dBm to linear scale according to
$P_{\mathrm{W}}(\alpha)
=10^{\frac{P(\alpha)-30}{10}}$.


\subsection{Utility and Regret}

Let $T(\alpha)$ denote the aggregate system throughput (Mbit/s) obtained when the transmit-power level $P(\alpha)$ is applied. To balance network performance and energy consumption, the arbiter evaluates each candidate action using the following utility function:

\begin{equation}
\label{eq:4}
U(\alpha)
=
T(\alpha)
-
w_{\mathrm{E}}
\,P_{\mathrm{W}}(\alpha),
\end{equation}
where $w_{\mathrm{E}}\ge0$ is an operator-defined energy-weight parameter that converts power consumption into the same utility scale as throughput. Larger values of $w_{\mathrm{E}}$ place greater emphasis on energy efficiency, whereas $w_{\mathrm{E}}=0$ reduces the objective to throughput maximization. The optimal action is defined as:
\begin{equation}
\label{eq:5}
\alpha^{\ast}
=\argmax_{\alpha \in \mathcal{A}}
U(\alpha),
\end{equation}
and the regret associated with selecting action $\alpha$ is
$r(\alpha)
=U(\alpha^{\ast})-U(\alpha)$. By construction, $r(\alpha)\ge0$, and $r(\alpha^{\ast})=0$. To compare performance across different values of $w_{\mathrm{E}}$, we also consider the following normalized regret:
\begin{equation}
\label{eq:6}
\bar r(\alpha)
=
\frac
{U(\alpha^{\ast})-U(\alpha)}
{U_{\max}-U_{\min}},
\end{equation}
where
\[
U_{\max}
=
\max_{\alpha \in \mathcal{A}} U(\alpha)
,\qquad
U_{\min}
=
\min_{\alpha \in \mathcal{A}} U(\alpha).
\]
A low regret indicates effective conflict resolution, with zero regret corresponding to the utility-maximizing arbitration decision.


\section{Problem Formulation}
\label{sec:problem}
The conflict-resolution problem considered in this paper is to select a single transmit-power action from the candidate set generated by the ES and CTO xApps. The objective is to maximize the energy-aware utility
defined in Section~\ref{sec:model}. Because the utility associated with
a candidate action cannot be observed before deployment, decisions must be made using estimates provided by an NDT. Hence, the resulting problem combines conflict resolution with
decision-making under model uncertainty, since discrepancies between the NDT and the live network may cause prediction accuracy to degrade over
time.

\subsection{Arbitration Objective}
Given conflicting transmit-power requests from the ES and CTO xApps, the arbiter must select a compromise action from the candidate set $\mathcal{A}$. The objective is to maximize the energy-aware utility introduced in Section~\ref{sec:model}. Formally, the arbitration problem is given by eq.(\ref{eq:5}). The action space in eq.(\ref{eq:2}) is already restricted to convex combinations of the ES and CTO requests, ensuring that the applied transmit power always lies within the interval $[P_{\mathrm{ES}},P_{\mathrm{CTO}}]$. No additional constraints are imposed at the arbitration layer. In particular, user scheduling, resource-block allocation, and lower-layer radio-resource-management functions remain internal to the underlying RAN stack. Likewise, QoS requirements are not explicitly enforced in the optimization objective and are considered only by the QoS-aware baseline methods used for comparison.

A key challenge is that the throughput function $T(\alpha)$ is not available in closed form. Instead, it is obtained from a detailed system-level simulator or from the live network itself. Consequently, the utility function in eq.(\ref{eq:4}) is effectively a black-box objective whose value can only be estimated through evaluation. This motivates the use of an NDT to predict the utility of candidate actions before deployment.

The optimal arbitration decision depends on the operator-defined energy weight $w_{\mathrm{E}}$. Denoting this dependency explicitly, $\alpha^{\ast}
=
\alpha^{\ast}(w_{\mathrm{E}})$. 
In the evaluation, the true optimum $\alpha^{\ast}$ is used only to compute regret metrics. During operation, the arbiter observes neither the complete utility landscape nor the optimal action.

\subsection{Twin-Assisted Decision Model}
To estimate the utility of candidate actions prior to execution, the arbiter queries an NDT. For each $\alpha \in \mathcal{A}$, the twin provides a predicted utility $\widehat{U}(\alpha)$. In contrast, the live network reveals only the realized utility of the action that is actually deployed,
$U_{\mathrm{live}}(\alpha_t)$.
Consequently, the twin is used to rank candidate actions, whereas the live network supplies feedback only for the selected action. 
However, the twin may become inaccurate over time because its internal representation of the network no longer matches reality. To model this effect in a controlled manner, we introduce a mismatch between the neighbor-cell transmit power assumed by the twin and the corresponding value in the live environment as follows:
\begin{equation}
\label{eq:9}
P_{\mathrm{neighbor}}^{(\mathrm{twin})}
=
P_{\mathrm{neighbor}}^{(\mathrm{live})}
+ d,
\end{equation}
where $d$ (in dB) denotes the drift magnitude, and all transmit-power values are expressed in dBm. When $d=0$, the twin and live environments are aligned and $\widehat{U}(\alpha)
=U_{\mathrm{live}}(\alpha)$ for all candidate actions. As $d$ increases, the prediction error grows, thus reducing the reliability of twin-based recommendations.

\subsection{Runtime Fidelity Estimation}
Since twin accuracy cannot be assumed a priori, the arbiter continuously monitors the discrepancy between predicted and realized utility. Let $\alpha_t$ denote the action selected at control step $t$. The instantaneous prediction error is
\begin{equation}
\label{eq:10a}
e_t
=
\left|
\widehat{U}(\alpha_t)
-
U_{\mathrm{live}}(\alpha_t)
\right|.
\end{equation}
To suppress short-term fluctuations and measurement noise, the arbiter maintains an exponentially weighted moving average (EWMA) of the prediction error as follows:
\begin{equation}
\label{eq:10b}
\varepsilon_t
=
\beta \varepsilon_{t-1}
+
(1-\beta)e_t,
\end{equation}
with initialization $\varepsilon_0=0$ and smoothing factor $\beta\in(0,1)$.
The quantity $\varepsilon_t$ serves as a runtime fidelity indicator. Small values suggest that the twin remains well aligned with the live network, whereas large values indicate increasing divergence and reduced confidence in twin-generated action recommendations.

\subsection{Fallback Action and Operating Assumptions}
In addition to the fidelity estimate, the arbiter maintains a live-network reference action that is independent of twin predictions. Specifically, it records the highest realized utility observed up to time $t$,

\begin{equation}
\label{eq:11}
U_{\mathrm{best},t}
=
\max_{k \le t}
U_{\mathrm{live}}(\alpha_k),
\end{equation}
and stores the corresponding action
\begin{equation}
\label{eq:12}
\alpha_{\mathrm{LKG},t}
=
\argmax_{\alpha_k,\,k\le t}
U_{\mathrm{live}}(\alpha_k),
\end{equation}
which we refer to as the \emph{last-known-good (LKG)} action. The rationale is that, if the twin becomes unreliable, the arbiter can fall back to an action that has previously demonstrated strong performance in the live network rather than continuing to rely on potentially inaccurate twin predictions.

The proposed framework is based on three assumptions as follows:
\begin{itemize}
\item \textbf{A1 (Calibrated warm-up):} During an initial warm-up period of $T_{\mathrm{w}}$ control steps, the twin is assumed to be well calibrated ($d=0$). Consequently, actions selected using the twin provide reliable live observations from which the LKG action can be established.

\item \textbf{A2 (Slowly varying optimum):} The utility-maximizing action evolves more slowly than the control timescale. Therefore, an action that recently performed well remains a meaningful fallback candidate. Under highly nonstationary network conditions, additional aging or exploration mechanisms may be required.

\item \textbf{A3 (Twin query budget):} The digital twin can evaluate all candidate actions in $\mathcal{A}$ within a single control interval, allowing the arbiter to rank actions before execution.
\end{itemize}

These elements define the information available to the arbiter and form the basis of the twin-fidelity-aware arbitration mechanism presented in the next section.

\section{Proposed Twin-Fidelity-Aware Arbitration Approach}
\label{sec:algo}
This section presents the proposed conflict-resolution mechanism for direct xApp conflicts over downlink transmit power. The key idea is to exploit the NDT whenever its predictions remain consistent with live-network observations while avoiding excessive reliance on the twin when its fidelity deteriorates. To achieve this objective, the proposed method combines three elements: (i) twin-based candidate-action evaluation, (ii) online fidelity monitoring based on prediction error, and (iii) a fallback mechanism that reuses the best utility-proven action previously observed in the live network.

\subsection{Twin-Preferred Action}
At each control step $t$, the NDT evaluates all candidate actions in $\mathcal{A}$ and predicts their utilities. The twin-preferred action is defined as
\begin{equation}
\label{eq:13}
\alpha_{\mathrm{twin},t}
=
\argmax_{\alpha \in \mathcal{A}}
\widehat{U}(\alpha).
\end{equation}
If the twin were perfectly accurate, repeatedly applying
$\alpha_{\mathrm{twin},t}$ would recover the utility-maximizing
decision. However, prediction errors may emerge as the twin drifts away
from the live network. Consequently, the twin recommendation is accepted
only when the estimated twin fidelity remains sufficiently high.

\subsection{Fidelity-Aware Hard Switching}
The proposed arbiter uses the EWMA fidelity signal defined in
Section~\ref{sec:problem} to determine whether the twin should be
trusted. Let $\tau$ denote a predefined fidelity threshold. When the
estimated mismatch remains below $\tau$, the twin recommendation is
applied. Otherwise, control is transferred to the last-known-good
(LKG) action learned from live observations. 
Formally, the applied action is as follows:
\begin{equation}
\label{eq:14}
\alpha_t
=
\alpha_{\mathrm{twin},t}
\mathbf{1}
\!\left[
\varepsilon_{t-1}<\tau
\right]
+
\alpha_{\mathrm{LKG},t-1}
\mathbf{1}
\!\left[
\varepsilon_{t-1}\ge\tau
\right],
\end{equation}
where $\mathbf{1}[\cdot]$ denotes the indicator function. 
Equivalently,
\[
\alpha_t=
\begin{cases}
\alpha_{\mathrm{twin},t},
&
\varepsilon_{t-1}<\tau,
\\[2mm]
\alpha_{\mathrm{LKG},t-1},
&
\varepsilon_{t-1}\ge\tau.
\end{cases}
\]
During the warm-up period, the arbiter always follows the twin and
collects live observations to initialize
$\alpha_{\mathrm{LKG}}$. The fidelity-based switching logic becomes
active only after a valid LKG action has been identified. 
Unlike soft-combination strategies that interpolate between candidate
actions, the proposed method performs a hard switch. This design choice
is motivated by the possibility of non-monotonic utility landscapes,
where an intermediate action may achieve lower utility than either of
the two endpoint actions. By switching directly to a previously
validated live-network action, the arbiter avoids traversing potentially
poor intermediate operating points.

\subsection{Online Learning of the Last-Known-Good Action}
The fallback action is continuously learned from live-network feedback. Specifically, the arbiter records the highest realized utility observed
so far, $U_{\rm best,t}$ (eq.(\ref{eq:11}))
and stores the corresponding action $\alpha_{\rm LKG,t}$ (eq.(\ref{eq:12})). The LKG mechanism provides a model-independent safety anchor. 

\begin{algorithm}[!t]
\caption{Fidelity-Aware Hard-Switching Arbiter}
\label{alg:arbiter}
\begin{algorithmic}[1]
\REQUIRE action grid $\mathcal{A} = \{0, 0.05, \dots, 1\}$; twin predictor $\widehat{U}(\cdot)$; fidelity threshold $\tau$; EWMA factor $\beta \in (0,1)$
\ENSURE sequence of applied actions $\{\alpha_{t}\}$
\STATE $\varepsilon \leftarrow 0$;\ \ $\alpha_{\mathrm{LKG}} \leftarrow \varnothing$;\ \ $U_{\mathrm{best}} \leftarrow -\infty$
\FOR{$t = 1, 2, \dots$}
    \STATE $\widehat{U}(\alpha) \leftarrow \mathrm{twin.predict}(\alpha)$ for all $\alpha \in \mathcal{A}$
    \STATE $\alpha_{\mathrm{twin}} \leftarrow \argmax_{\alpha \in \mathcal{A}} \widehat{U}(\alpha)$
    \IF{$\varepsilon < \tau$ \OR $\alpha_{\mathrm{LKG}} = \varnothing$}
        \STATE $\alpha_{t} \leftarrow \alpha_{\mathrm{twin}}$
    \ELSE
        \STATE $\alpha_{t} \leftarrow \alpha_{\mathrm{LKG}}$
    \ENDIF
    \STATE Apply $P(\alpha_{t})$; observe live utility $U_{\mathrm{live}}(\alpha_{t})$
    \STATE $e \leftarrow |\widehat{U}(\alpha_{t}) - U_{\mathrm{live}}(\alpha_{t})|$
    \STATE $\varepsilon \leftarrow \beta\,\varepsilon + (1 - \beta)\,e$
    \IF{$U_{\mathrm{live}}(\alpha_{t}) > U_{\mathrm{best}}$}
        \STATE $U_{\mathrm{best}} \leftarrow U_{\mathrm{live}}(\alpha_{t})$;\ \ $\alpha_{\mathrm{LKG}} \leftarrow \alpha_{t}$
    \ENDIF
\ENDFOR
\end{algorithmic}
\end{algorithm}

Algorithm \ref{alg:arbiter} summarizes the operation of the proposed arbitration solution.

\subsection{Theoretical Properties}
\subsubsection{Detection Delay}
A sustained prediction mismatch produces a predictable growth of the EWMA fidelity signal. Assuming a constant mismatch $e_t=m>\tau$ and an initially calibrated twin ($\varepsilon_0=0$), EWMA evolves as $\varepsilon_t =m(1-\beta^t)$. Consequently, the fidelity threshold is crossed after:
\begin{equation}
\label{eq:15}
t_{\mathrm{detect}}
=
\left\lceil
\frac{\ln(1-\tau/m)}
{\ln \beta}
\right\rceil \; \text{ control steps. }
\end{equation}
The detection delay decreases as the sustained prediction mismatch
increases and increases as the smoothing factor $\beta$ approaches one.

\subsubsection{Reversibility}
The proposed mechanism is inherently reversible. The fidelity estimate
continues to be updated while the fallback action is active. Therefore,
if the twin subsequently realigns with the live network, the prediction
error decreases, the EWMA decays below the threshold, and control
automatically returns to the twin-selected action.

\subsubsection{On-policy fidelity}
Since prediction errors are evaluated only on actions that are
actually executed, the resulting fidelity estimate is on-policy. Hence, the monitor measures the accuracy of the twin along the realized operating trajectory without requiring additional exploratory probing of
the action space.

\subsubsection{Complexity}
At each step, the arbiter queries the twin $|\mathcal{A}|$ times (here, 21 queries), computes the argmax, performs one threshold check, applies the action, and updates two scalar variables ($\varepsilon$ and $U_{\mathrm{best}}$). The per-step computational complexity is $\mathcal{O}(|\mathcal{A}|)$, which is minimal and identical to that of blind twin selection. The storage complexity is $\mathcal{O}(1)$ beyond the twin itself, requiring only the scalars $\varepsilon$, $\alpha_{\mathrm{LKG}}$, and $U_{\mathrm{best}}$.

\section{Evaluation Methodology}
\label{sec:eval}
This section describes the simulation environment, the NDT emulation procedure, the arbitration methods compared, and the performance metrics used to evaluate the proposed approach. The objective is to quantify the impact of twin drift on conflict-arbitration decisions and to assess whether online fidelity monitoring improves robustness relative to conventional twin-based and QoS-driven conflict-resolution strategies.

\begin{figure}[!t]
\centering
\includegraphics[width=\columnwidth]{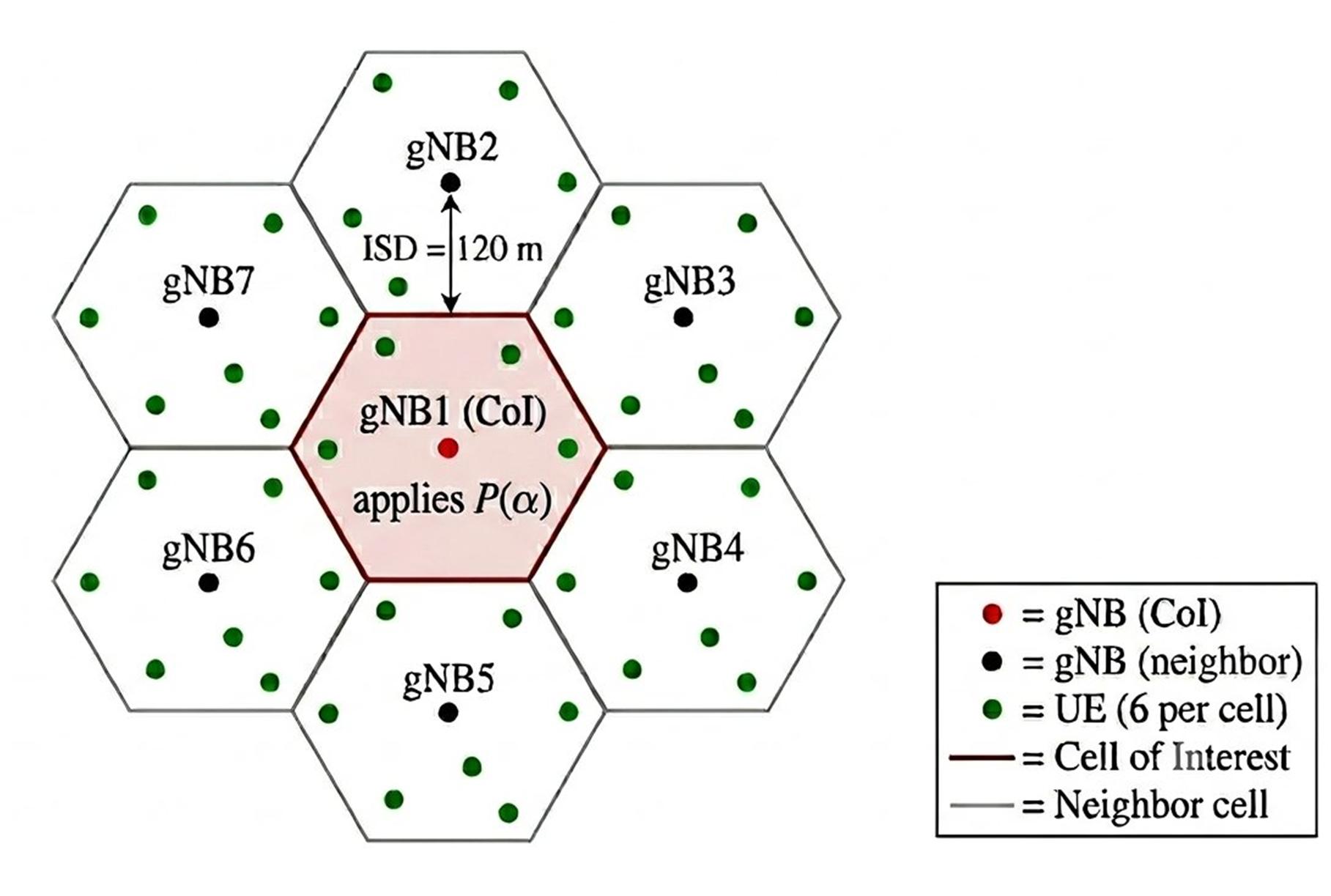}
\caption{Seven-cell hexagonal evaluation scenario. The cell of interest (gNB1) applies the arbitrated downlink transmit power $P(\alpha)$, while the six neighboring cells act as fixed-power interferers. Parameters are listed in Table~\ref{tab:params}.}
\label{fig:scenario}
\end{figure}

\begin{table}[!t]
\caption{Simulation Parameters}
\label{tab:params}
\centering
\renewcommand{\arraystretch}{1.2}
\begin{tabular}{p{0.40\columnwidth} p{0.50\columnwidth}}
\hline
\textbf{Parameter} & \textbf{Value}\\
\hline
Cell layout & Seven-cell hexagonal (one cell of interest, six neighbors)\\
Inter-site distance / cell radius & 120~m / 60~m\\
Carrier / bandwidth / subcarrier spacing & 2.5~GHz / 10~MHz / 30~kHz\\
gNB / UE heights & 30~m / 1.5~m\\
Users and placement & 6 UEs per cell (42 total), uniform angle, radius 0.75--0.95 of cell radius\\
Scheduler / traffic & Round-robin; always-on downlink source per UE (saturating offered load, 1500-byte packets)\\
CoI transmit-power range $[P_{\mathrm{ES}}, P_{\mathrm{CTO}}]$ & $[10, 46]$~dBm\\
Neighbor transmit power (live) & 32~dBm (fixed)\\
Action grid $\mathcal{A}$ & $\{0, 0.05, \dots, 1\}$ (21 actions)\\
Scenario length per $(\alpha, d)$ sample & 50 frames (0.5~s)\\
Control steps per run (warm-up + main) & $T = 40$ (10 calibrated warm-up + 30 main)\\
EWMA factor $\beta$ & 0.70\\
Soft-variant trust decay $\mu$ & 0.30\\
Fidelity threshold $\tau$ & 1 (selected from $\{1,2,3,4,5\}$; see Sec.~\ref{subsec:config})\\
Operator energy weights $w_{\mathrm{E}}$ (priority study) & $\{0.05, 0.1, 0.2, 0.5, 1, 2, 5\}$\\
Operator energy weights (drift study) & $\{0.1, 0.2, 1\}$\\
Twin drift levels $d$ & $\{0, 2, 4, 6, 8, 10\}$~dB\\
QoS targets & Throughput floor: 70\% of per-seed max aggregate throughput; power ceiling: $P(0.25) = 19$~dBm ($\approx 0.079$~W)\\
Independent random seeds & 20\\
\hline
\end{tabular}
\end{table}

\subsection{Evaluation Scenario}
\label{subsec:scenario}

The evaluation scenario, shown in Fig.~\ref{fig:scenario}, consists of a seven-cell hexagonal deployment in which the central cell of interest (gNB1) hosts the conflicting ES and CTO xApps. The conflict-resolution mechanism determines the downlink transmit power of gNB1 within the range $[10,46]$~dBm, whereas the six neighboring cells (gNB2--gNB7) operate at a fixed transmit power of 32~dBm. 
The scenario is implemented using the MATLAB 5G Toolbox system-level simulator based on \texttt{wirelessNetworkSimulator}, \texttt{nrGNB}, and \texttt{nrUE} objects. The network operates at a carrier frequency of 2.5~GHz with 10~MHz bandwidth and 30~kHz subcarrier spacing. Downlink scheduling uses the default round-robin scheduler.

Each cell serves six UEs, resulting in 42 UEs across the deployment. UEs are dropped uniformly in angle and placed between 75\% and 95\% of the cell radius, thus emphasizing cell-edge conditions where inter-cell interference is most pronounced. An always-on traffic source generates 1500-byte downlink packets with an offered load of 100~Mbit/s per UE, creating a saturated operating regime in which throughput is primarily limited by radio conditions and interference. Unless otherwise stated, aggregate throughput is measured across all seven cells. Consequently, the utility function of eq.\eqref{eq:4} captures both the throughput gains and the interference costs associated with a transmit-power decision.

\subsection{Digital-Twin Emulation and Experimental Procedure}
\begin{figure*}[!t]
\centering
\includegraphics[width=0.7\linewidth]{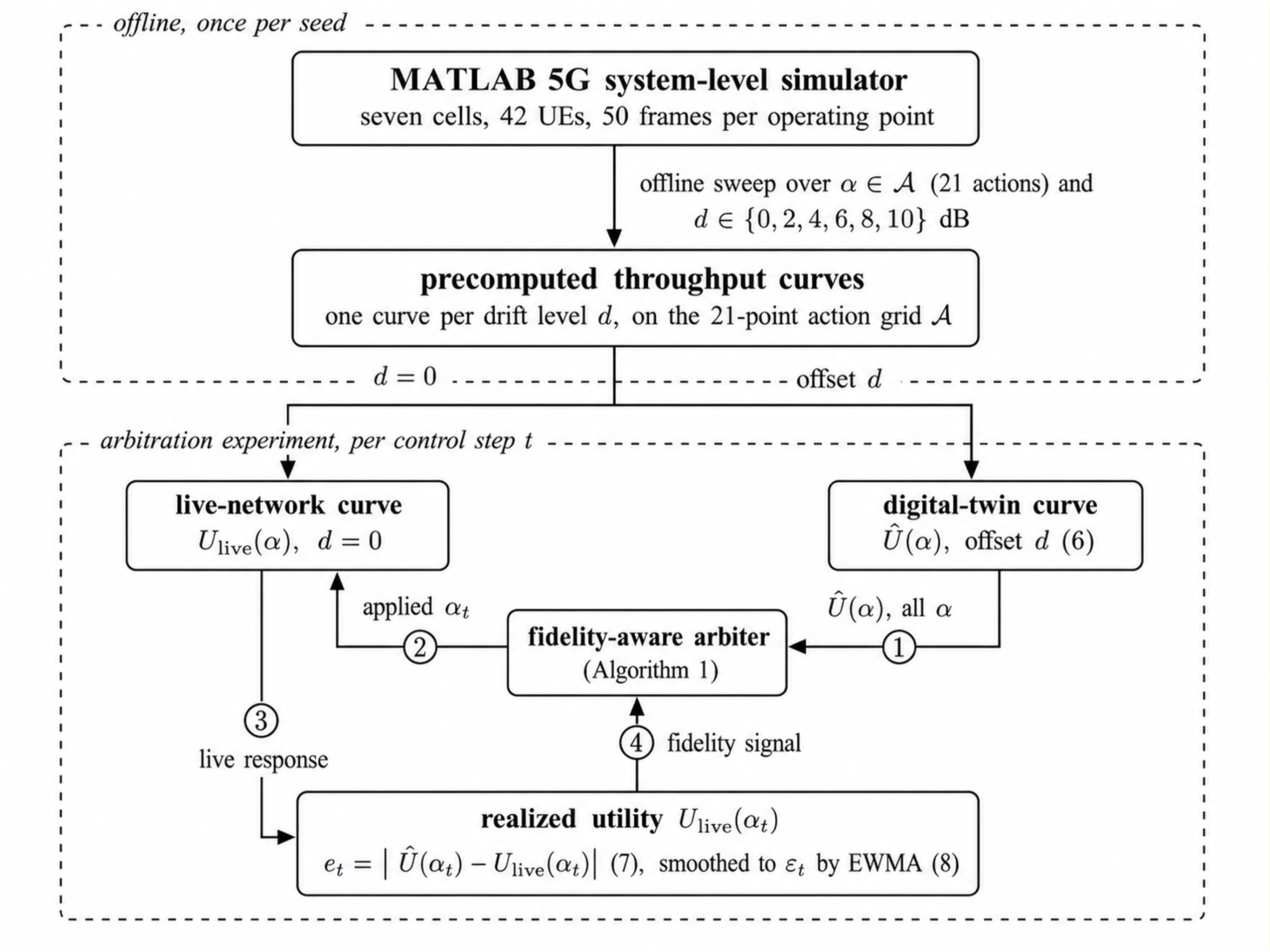}
\caption{Evaluation data flow. For each seed, one system-level model
precomputes the live-network and NDT throughput curves. These curves are
then used to evaluate the fidelity-aware arbiter introduced in
Fig.~\ref{fig:arch}. Steps 1--4 execute at every control step.}
\label{fig:dataflow}
\end{figure*}

Both the live network and the NDT are obtained from the same MATLAB system-level model, evaluated under two different assumptions about the neighboring cells. The twin's assumed neighbor transmit power is offset by $d$~dB from the live value, following eq.(\ref{eq:9}). Fig.~\ref{fig:dataflow} shows the resulting data flow. When $d=0$, the twin reproduces the live environment exactly and the predicted and realized utilities coincide. Increasing $d$ progressively degrades twin fidelity by biasing the interference conditions assumed by the twin. Deriving both sides from one validated model makes the twin--live discrepancy exactly the injected offset $d$, so the arbiter's behavior is attributable to the fidelity monitor rather than to differences between two simulator implementations. Accordingly, the NDT is homogeneous with the live model.

For each random seed, the aggregate-throughput response of the live network is precomputed on the 21-point action grid, each operating point simulated for 50 frames (0.5~s), and the corresponding twin throughput curves are generated for each drift level. Utility values during arbitration are read from these curves, with linear interpolation where required. Each experiment runs $T = 40$ control steps, comprising a 10-step warm-up at $d=0$, consistent with Assumption A1, followed by a 30-step evaluation phase over which all reported metrics are computed.

\subsection{Compared Methods}

The proposed twin-fidelity-aware arbiter is compared against the following baselines:
\begin{itemize}
\item \textbf{ES-only:}
Always applies the ES request ($\alpha=0$).
\item \textbf{CTO-only:}
Always applies the CTO request ($\alpha=1$).
\item \textbf{Naive Blend:}
Applies the fixed midpoint action $\alpha=0.5$.
\item \textbf{COMIX-style Selection}~\cite{ref5}:
Always applies the twin-preferred action
\(
\argmax_{\alpha}\widehat U(\alpha)
\)
without fidelity monitoring or fallback. This corresponds to the limiting case $\tau\rightarrow\infty$ of the proposed framework.
\item \textbf{QACM-style Selection}~\cite{ref8}:
Selects the action that maximizes QoS-threshold satisfaction based on
throughput and power objectives, with the throughput target evaluated on
the twin-predicted curve. Unlike the proposed method, it neither optimizes
the utility function of eq.\eqref{eq:4} nor explicitly accounts for twin
fidelity.
\item \textbf{Soft-ES:} A fidelity-aware soft arbitration rule that scales
the twin-selected action by a trust factor that decays with the fidelity
error. Let $\mu>0$ be a decay parameter controlling how rapidly trust
decreases as the twin-fidelity error $\varepsilon$ increases (unless
otherwise stated, $\mu=0.30$). The trust factor and applied action are
$q_t=\exp(-\mu\varepsilon_{t-1})$ and
$\alpha_t=\Pi_{[0,1]}\!\left(q_t\,\alpha_{\mathrm{twin},t}\right)$,
where $\varepsilon_{t-1}$ is the variant's own EWMA fidelity estimate,
updated as in eq.\eqref{eq:10b} and available before acting, and
$\Pi_{[0,1]}(\cdot)$ denotes clipping to $[0,1]$. As twin fidelity
deteriorates, the operating point moves progressively toward the ES action.


\item \textbf{Soft-LKG:}
A fidelity-aware interpolation rule that blends the twin-preferred action
with the last-known-good action using the same trust factor
$q_t=\exp(-\mu\varepsilon_{t-1})$. The applied action is
$\alpha_t=\Pi_{[0,1]}\!\left[q_t\,\alpha_{\mathrm{twin},t}
+\left(1-q_t\right)\alpha_{\mathrm{LKG},t-1}\right]$,
so that the operating point moves toward the last-known-good action as
twin fidelity deteriorates.
\end{itemize}

\subsection{Parameter Configuration}
\label{subsec:config}

Unless otherwise specified, the proposed arbiter operates with EWMA factor $\beta=0.70$ and fidelity threshold $\tau=1$. 
The choice of $\beta$ assigns a weight of $0.30$ to the most recent observation while providing moderate smoothing of short-term fluctuations. This value is fixed throughout all experiments and is not tuned. Subsection~\ref{subsec:betasens} below evaluates sensitivity to this choice. The threshold $\tau$ is selected from the candidate set
$\{1,2,3,4,5\}$
using mean regret over all evaluation runs. Among all tested values, $\tau=1$ consistently produced the lowest average regret both in the operator-priority study and in the twin-drift experiments.
All simulation and algorithm parameters are summarized in Table~\ref{tab:params}.

\subsection{Performance Metrics and Statistical Methodology}
\label{subsec:metrics}

Performance is evaluated using two complementary metrics.

\textbf{Normalized Regret:}
The primary metric is the normalized regret defined in eq.\eqref{eq:6}. This metric quantifies the utility loss incurred by an arbitration decision relative to the utility-maximizing action and enables fair comparison across different operator priorities.
Regret is reported in normalized form per \eqref{eq:6} when comparing across
operator priorities, and in unnormalized form (in the units of the utility
\eqref{eq:4}) within a fixed priority in the drift experiments.

\textbf{QoS-Satisfaction Ratio:}
The second metric measures the fraction of QoS objectives satisfied by the selected action. Two QoS targets are considered: (i) a throughput floor equal to 70\% of the per-seed maximum aggregate throughput, and (ii) a power ceiling corresponding to the transmit power associated with $\alpha=0.25$ (19~dBm).

In the operator-priority study, each metric is first averaged, within every
seed, over the 30 main-phase steps, the six drift levels, and the seven
operator energy weights, yielding one value per independent seed. The
reported means and Student-$t$ 95\% confidence intervals are then computed
across the 20 independent seeds, so that correlated within-seed samples are
not treated as independent observations. 
For the twin-drift study, results are reported separately for each drift level and operator weight. Confidence intervals are again computed using the distribution over the 20 independent random seeds. Finally, claims regarding robustness relative to the COMIX-style baseline are based on paired per-seed regret differences and are reported only when the corresponding 95\% confidence interval excludes zero.

\section{Results and Discussion}
\label{sec:results}

This section evaluates the proposed twin-fidelity-aware arbitration mechanism from complementary perspectives. First, we examine how the utility-maximizing conflict-resolution decision varies with the operator's energy-efficiency preference. Second, we evaluate the robustness of the proposed method under increasing digital-twin drift and compare it with twin-based, QoS-oriented, and fixed-action baselines. Third, we investigate QoS satisfaction and parameter sensitivity to better understand the trade-offs introduced by fidelity-aware arbitration. Finally, we assess sensitivity to the EWMA smoothing
factor. Unless otherwise stated, all results are obtained using the seven-cell scenario described in Section~\ref{sec:eval}, and reported values correspond to sample means with 95\% confidence intervals (CI).

\subsection{Impact of Operator Priorities}
\subsubsection{Utility Landscape and Optimal Operating Point}

\begin{figure*}[!t]
\centering
\includegraphics[width=\linewidth]{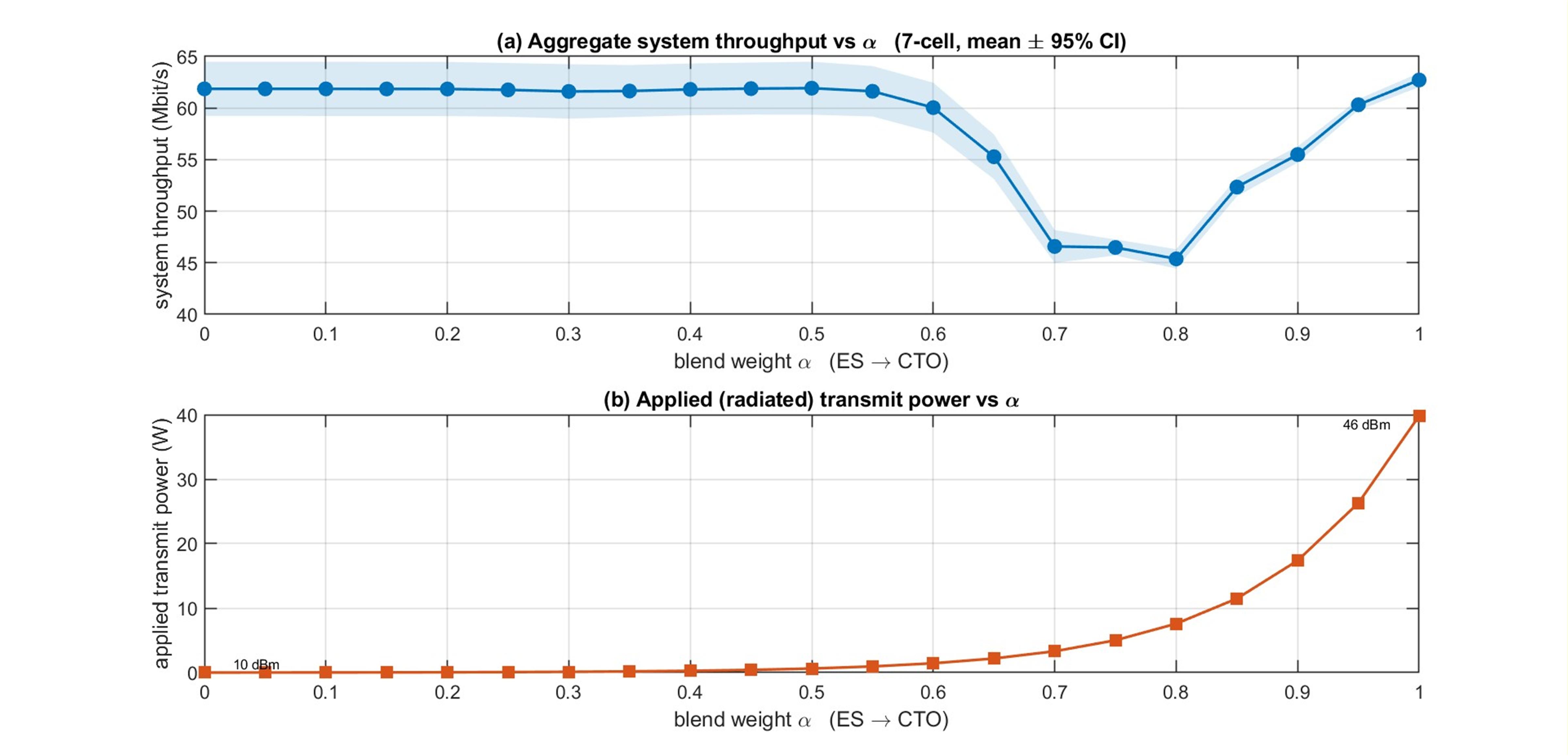}
\caption{Measured operating characteristics of the seven-cell scenario versus the blend weight $\alpha$: (a) aggregate system throughput (Mbit/s, mean $\pm$ 95\% CI); (b) applied radiated transmit power (W), with endpoint values in dBm.}
\label{fig:tradeoff}
\end{figure*}

Fig.~\ref{fig:tradeoff} illustrates the operating characteristics of the considered seven-cell scenario as a function of the blend weight $\alpha$. Increasing $\alpha$ moves the applied transmit power from the ES request toward the CTO request. Aggregate system throughput remains approximately constant at low $\alpha$, decreases within the interference-sensitive region around $\alpha \approx 0.7$--$0.8$, and partially recovers near the CTO endpoint. In contrast, transmit power increases exponentially with $\alpha$. 

Hence, the resulting optimization problem is nontrivial. Indeed, increasing transmit power may improve throughput in some regions, but it simultaneously increases energy consumption and may intensify inter-cell interference. Consequently, neither the ES nor CTO endpoint is universally optimal.

\begin{figure}[!t]
\centering
\includegraphics[trim={25 0 30 0},clip,width=\columnwidth]{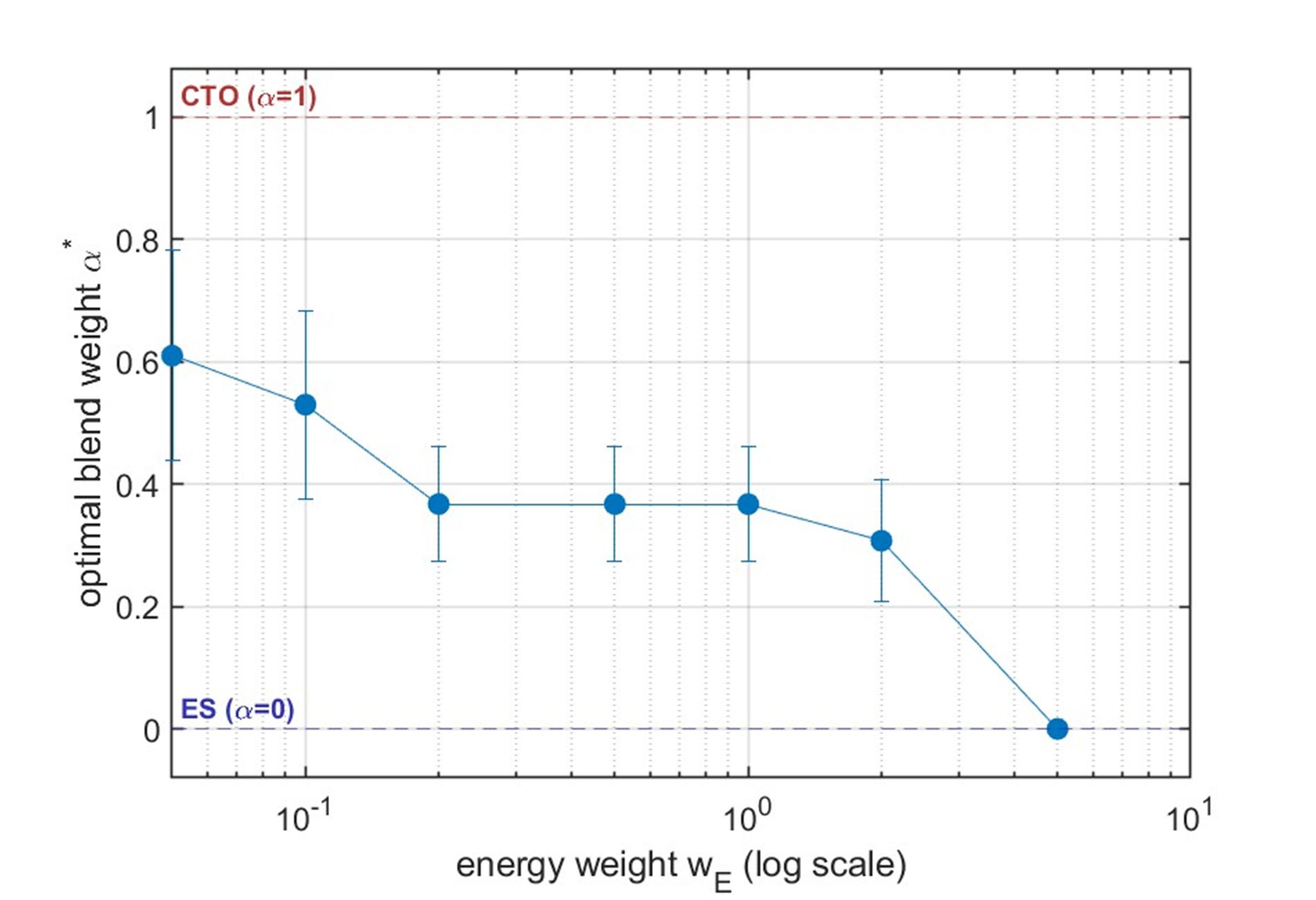}
\caption{Optimal blend weight $\alpha^{\ast}$ versus operator energy weight $w_{\mathrm{E}}$ in the seven-cell scenario (mean $\pm$ 95\% CI).}
\label{fig:alphastar}
\end{figure}

Fig.~\ref{fig:alphastar} shows the optimal blend weight $\alpha^\ast$ as a function of the operator energy weight $w_{\mathrm E}$. When energy is lightly penalized ($w_{\mathrm E}=0.05$ or $0.1$), the optimum lies in the interior of the action space, corresponding to a throughput-oriented operating point. As $w_{\mathrm E}$ increases, the optimum progressively shifts toward lower transmit power and eventually converges to the ES boundary $\alpha=0$ at $w_{\mathrm E}=5$. These results demonstrate that the preferred conflict-resolution action depends strongly on operator priorities and motivate adaptive arbitration rather than static policies such as ES-only, CTO-only, or a fixed midpoint.

\subsubsection{Utility-Regret Performance}

Table~\ref{tab:regret} reports the normalized utility regret averaged across all evaluated energy-priority settings. The proposed hard-switching arbiter achieves the lowest regret of
$0.017 \pm 0.006$, thus outperforming all competing methods. Among the
baselines, Soft-ES provides the closest performance with a regret of
$0.036 \pm 0.013$, whereas the COMIX-style selector reaches
$0.159 \pm 0.052$. Static policies perform substantially worse,
particularly CTO-only, whose operation in the high-power region results
in a regret of $0.665 \pm 0.037$.


\begin{table}[!t]
\caption{Normalized Utility Regret Across Operator Priorities}
\label{tab:regret}
\centering
\renewcommand{\arraystretch}{1.2}
\begin{tabular}{lc}
\hline
\textbf{Method} & \textbf{Normalized Regret}\\
\hline
Proposed hard-switch ($\tau = 1$) & $0.017 \pm 0.006$\\
Soft-ES & $0.036 \pm 0.013$\\
ES-only & $0.046 \pm 0.015$\\
QACM-style & $0.046 \pm 0.015$\\
Naive blend & $0.053 \pm 0.018$\\
Soft-LKG & $0.059 \pm 0.017$\\
COMIX-style & $0.159 \pm 0.052$\\
CTO-only & $0.665 \pm 0.037$\\
\hline
\end{tabular}
\end{table}

The superiority of the proposed approach stems from its combination of twin-assisted optimization and online reliability monitoring. Unlike blind twin-based arbitration, the proposed method does not continue to follow NDT recommendations once evidence of prediction drift emerges.

\subsection{Impact of Digital-Twin Drift}
\subsubsection{Drift Sensitivity of Twin-Based Arbitration}
\begin{figure}[!t]
\centering
\includegraphics[trim={30 0 30 0},clip,width=\columnwidth]{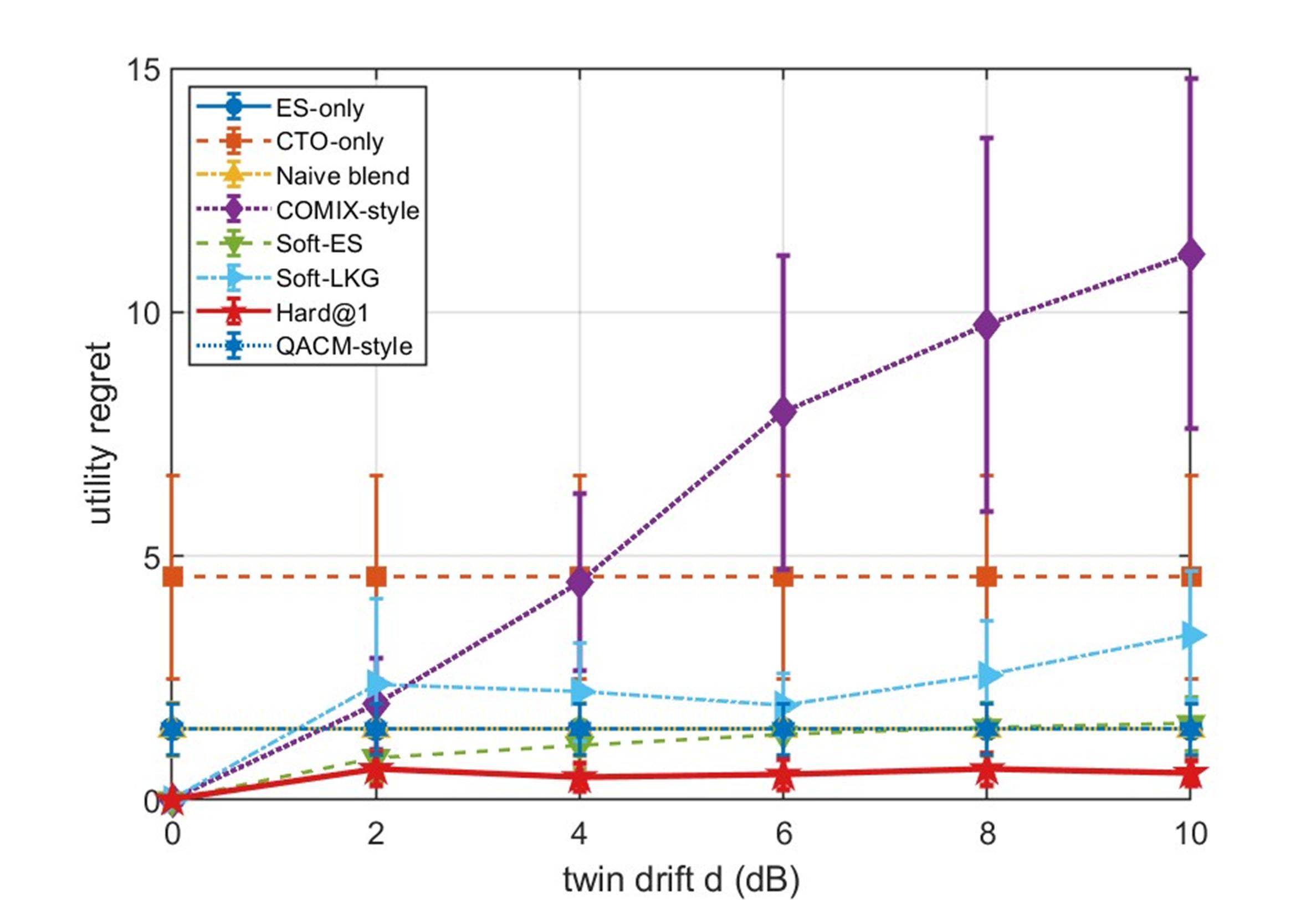}
\caption{Utility regret versus digital-twin drift at an interior optimum
($w_{\mathrm{E}} = 0.1$, mean $\pm$ 95\% CI).}
\label{fig:regretdrift}
\end{figure}

\begin{figure}[!t]
\centering
\includegraphics[trim={30 0 30 0},clip,width=\columnwidth]{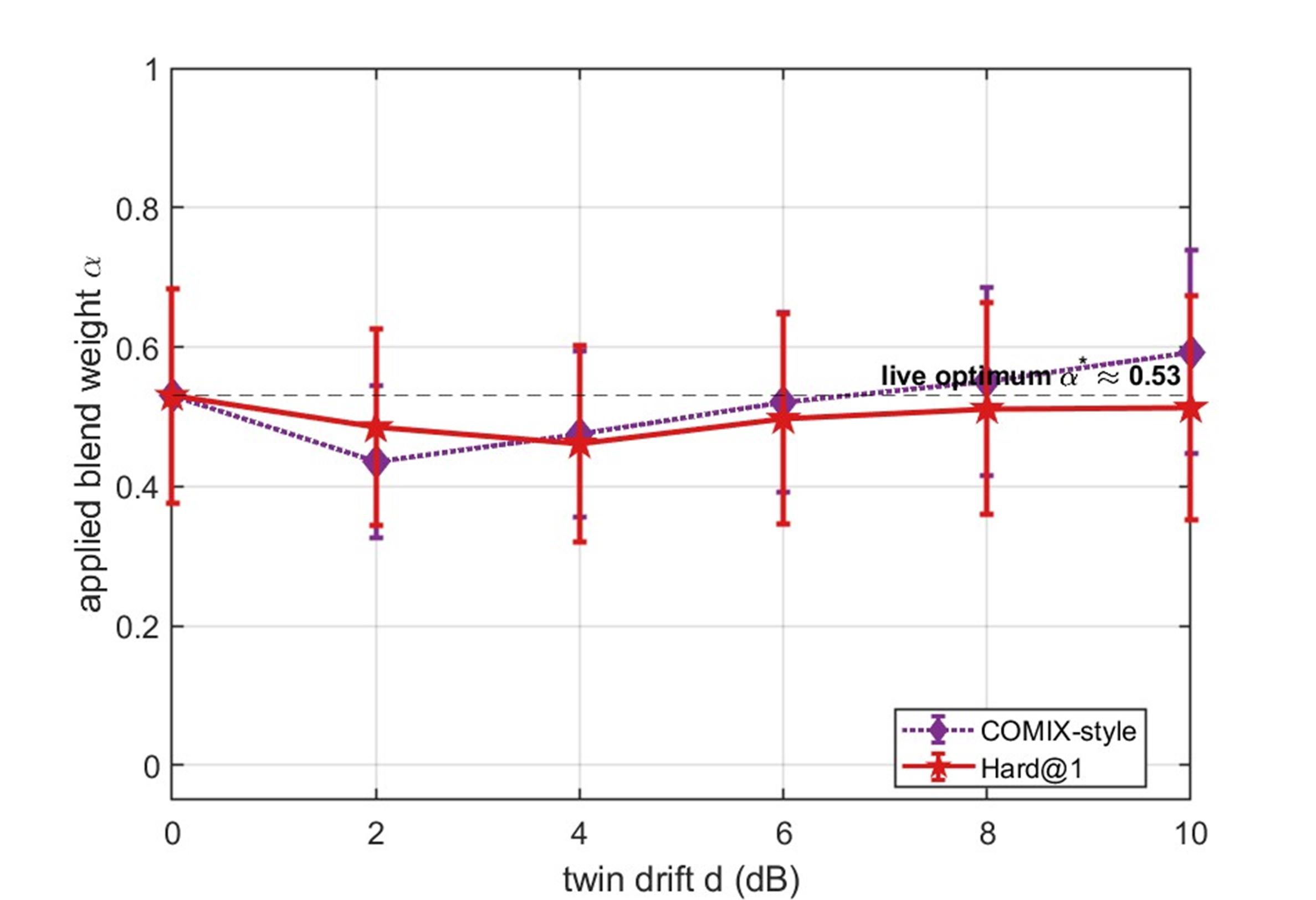}
\caption{Applied blend weight $\alpha$ versus digital-twin drift
($w_{\mathrm{E}} = 0.1$, mean $\pm$ 95\% CI).}
\label{fig:alphadrift}
\end{figure}

To isolate the role of twin fidelity, we first consider the representative
case $w_{\mathrm{E}} = 0.1$, where the live optimum lies in the interior of
the action space ($\alpha^\ast \approx 0.53$). Fig.~\ref{fig:regretdrift}
presents the utility regret of all compared methods under increasing twin
drift. The static baselines (ES-only, CTO-only, and naive blend) are insensitive to drift by construction, since their actions never consult the NDT. The QACM-style baseline evaluates its throughput target on the twin-predicted curve and therefore consults the NDT. However, under the selected QoS targets, its QoS-optimal action remains at the low-power operating point across all drift levels, so it appears drift-insensitive in these experiments. Thus, ES-only and QACM-style arbitration operate near the low-power region and incur a moderate but constant regret, as neither exploits the interior optimum. The naive blend remains constant at an intermediate level, while CTO-only is the worst static policy as it operates in the high-power, interference-limited region. The utility-driven twin-following methods separate as drift grows. The COMIX-style selector achieves zero regret at $d = 0$, where the twin and live networks coincide, but its regret grows rapidly with
drift, reaching $11.19 \pm 3.58$ at $10$~dB, because it continues to follow the biased twin predictions. The soft variants degrade more smoothly. Indeed, Soft-ES contracts the applied action toward the ES endpoint as the fidelity error grows, which steers its loss toward the ES-only level, whereas Soft-LKG interpolates between the twin-preferred and LKG actions and can hence traverse lower-utility regions of the non-monotonic utility landscape. The proposed hard-switching arbiter, denoted ``Hard@1'' ($\tau=1$) in Figs.~\ref{fig:regretdrift}, \ref{fig:alphadrift},
\ref{fig:throughput}, and~\ref{fig:power}, 
maintains regret below $0.65$ across the examined drift range, whereas the regret of blind twin-based selection increases sharply with drift.

Fig.~\ref{fig:alphadrift} reveals the underlying mechanism of the
twin-following methods. As twin predictions become increasingly biased, the
COMIX-style action progressively moves away from the live-optimal operating
region, whereas the proposed method remains anchored near the live-validated
operating point. Although the displacement in $\alpha$ appears modest, it
occurs on a steep section of the utility landscape, causing substantial
regret growth. Similar behavior is observed for $w_{\mathrm{E}} = 0.2$,
confirming that the phenomenon is not specific to a single energy-weight
configuration.

\subsubsection{Robustness of the Proposed Method}\label{sssec:robustness}

As shown in Fig.~\ref{fig:regretdrift}, the proposed method exhibits
fundamentally different behavior. Once the fidelity estimate exceeds the
threshold, control shifts from the twin-preferred action to the
live-validated last-known-good action, preventing sustained operation in regions favored only by a degraded twin. At $w_{\mathrm{E}} = 0.1$, the proposed method maintains regret below $0.65$ across all drift levels and achieves only $0.55 \pm 0.25$ regret even at $10$~dB drift. In contrast, the COMIX-style selector reaches $11.19 \pm 3.58$
(Fig.~\ref{fig:regretdrift}), a more-than-twentyfold difference in favor of
the proposed arbiter. 

\begin{figure}[!t]
\centering
\includegraphics[trim={30 0 30 0},clip,width=\columnwidth]{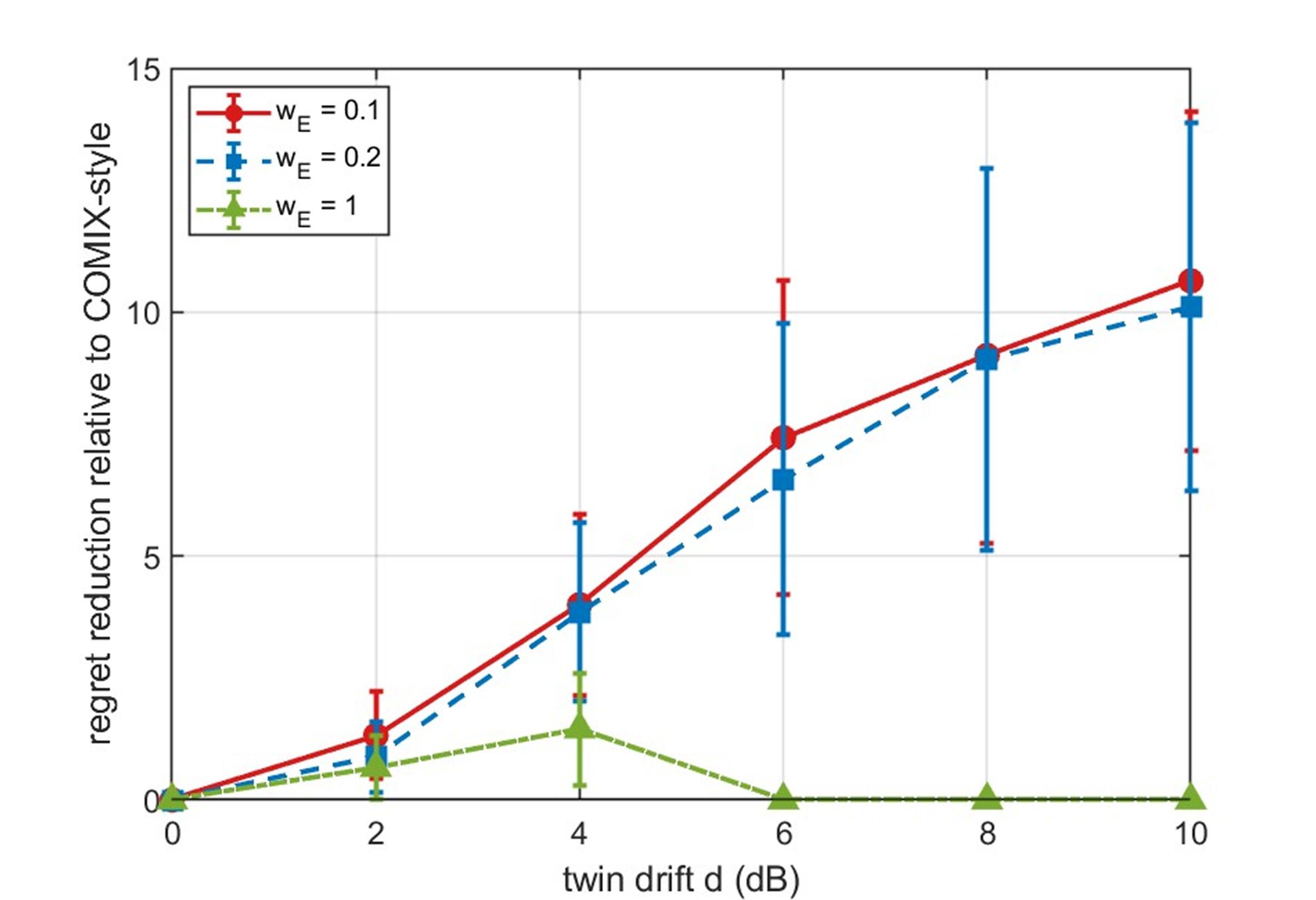}
\caption{Utility-regret reduction of the proposed hard-switching arbiter relative to COMIX-style selection versus digital-twin drift, for $w_{\mathrm{E}} \in \{0.1,0.2,1\}$.}
\label{fig:reduction}
\end{figure}

Fig.~\ref{fig:reduction} summarizes the paired per-seed regret reduction of the proposed method relative to the COMIX-style selector. In the interior-optimum regimes $w_{\mathrm{E}} = 0.1$ and $w_{\mathrm{E}} = 0.2$, the reduction is positive at every nonzero drift level and grows with drift. At $w_{\mathrm{E}} = 0.1$, it increases from $+1.32 \pm 0.90$ at $2$~dB to $+10.64 \pm 3.47$ at $10$~dB, with the $95\%$ CI excluding zero throughout, indicating statistically significant robustness gains. For
$w_{\mathrm{E}} = 1$, the gain is smaller because the biased twin moves
toward a low-power region that is already close to the live optimum; this
boundary-optimum regime is examined in Subsection~\ref{sssec:boundary}.
Fidelity monitoring thus provides the largest benefit when drift displaces the twin-selected action away from an interior optimum.

\begin{figure}[!t]
\centering
\includegraphics[trim={30 0 30 0},clip,width=\columnwidth]{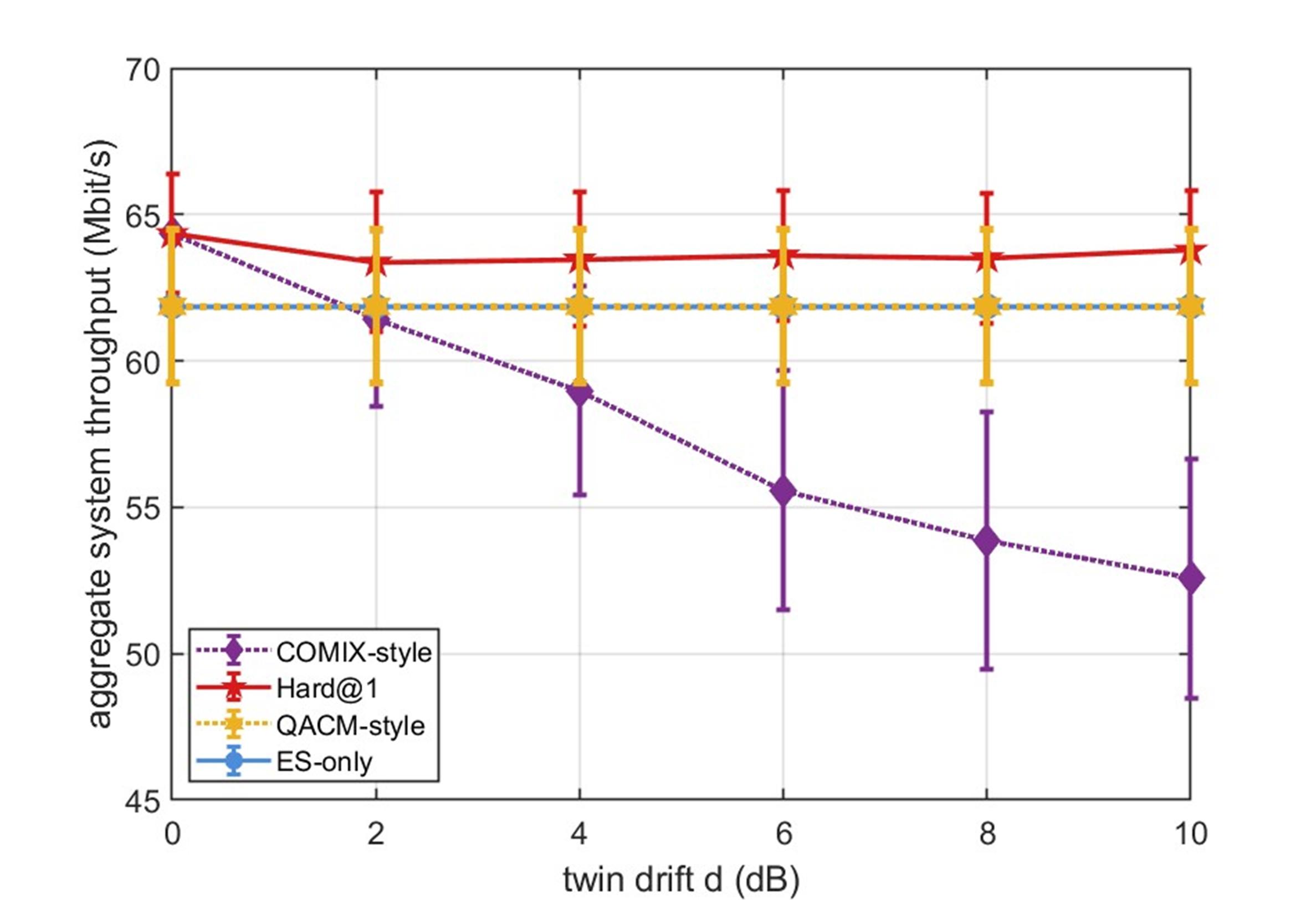}
\caption{Achieved aggregate system throughput at the applied operating point versus digital-twin drift ($w_{\mathrm E}=0.1$, mean $\pm$ 95\% CI).}
\label{fig:throughput}
\end{figure}

\begin{figure}[!t]
\centering
\includegraphics[trim={30 0 30 0},clip,width=\columnwidth]{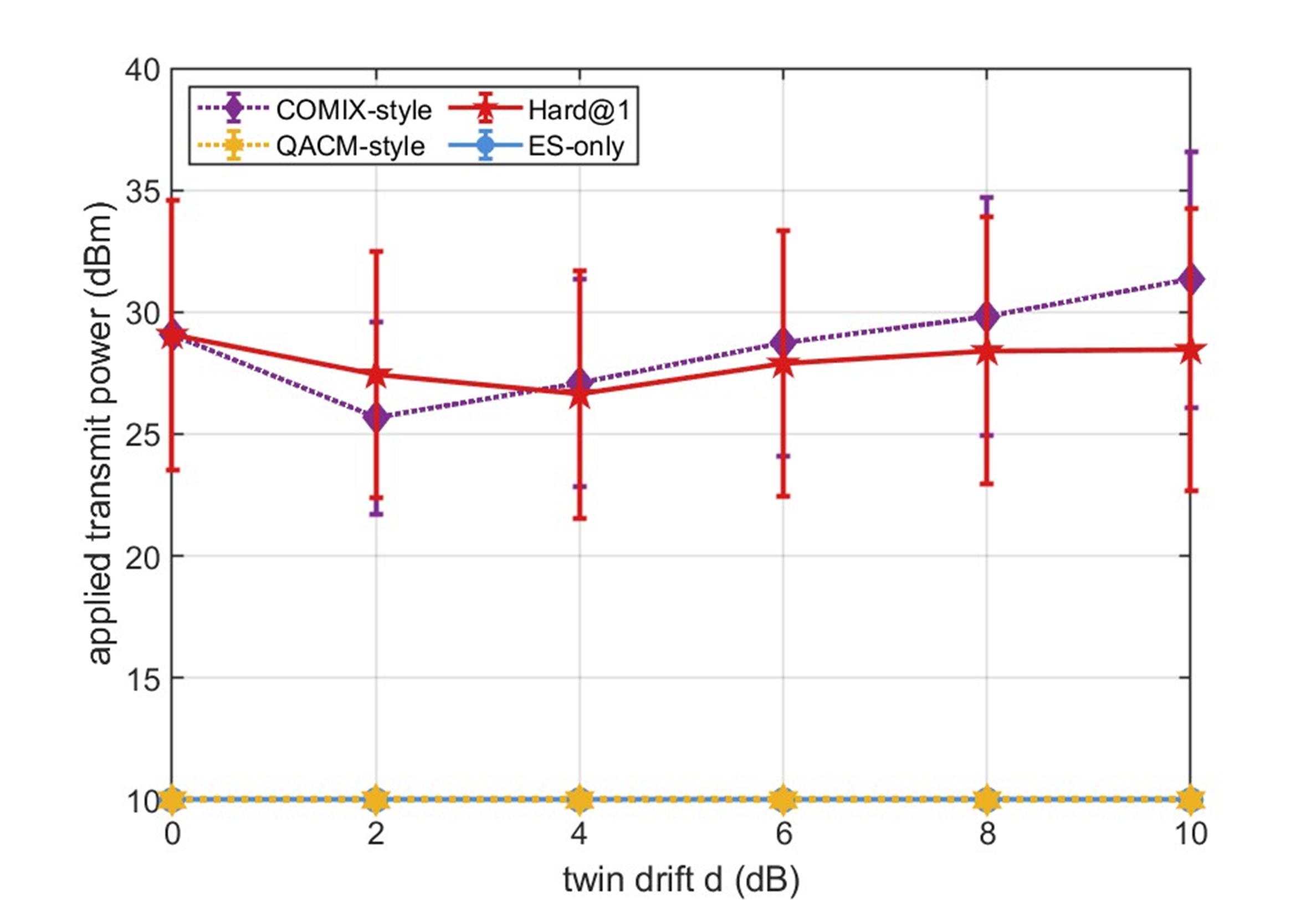}
\caption{Applied radiated transmit power versus digital-twin drift ($w_{\mathrm E}=0.1$, mean $\pm$ 95\% CI).}
\label{fig:power}
\end{figure}

The same trend is reflected in operational performance. At 10~dB drift, the proposed method maintains an aggregate throughput of approximately 64~Mbit/s, whereas the COMIX-style selector falls to approximately 53~Mbit/s (Fig.~\ref{fig:throughput}). Likewise, the proposed method maintains a stable power operating point near 28--29~dBm despite substantial twin degradation (Fig.~\ref{fig:power}).

\subsubsection{Boundary-Optimum Scenario}\label{sssec:boundary}

The largest benefits of fidelity monitoring occur when digital-twin drift displaces the operating point away from an interior optimum. A different behavior emerges when the live optimum lies near a boundary of the action space.

For $w_{\mathrm E}=1$, the optimum shifts toward the low-power region ($\alpha^\ast \approx 0.37$). Under increasing drift, the COMIX-style selector naturally moves toward the ES boundary. In this particular regime, the twin bias happens to guide the action toward a region that remains close to the live optimum. Consequently, COMIX-style selection, the proposed method, the soft variants, and ES-only converge to similar performance at high drift levels.

Importantly, however, the proposed method does not incur a penalty in this regime. When the twin continues to recommend a near-optimal action, fidelity monitoring does not force a harmful intervention, allowing the proposed method to match the strongest baselines. This result delineates the scope of the gain provided by fidelity monitoring while demonstrating that the safeguard does not degrade performance when intervention is unnecessary.

\subsection{QoS-Oriented Perspective}

Table~\ref{tab:qos} evaluates performance from a QoS-oriented perspective. As expected, ES-only and QACM-style arbitration achieve the highest QoS-satisfaction ratio because the selected QoS targets strongly favor low-power operating points. The proposed method, Soft-ES, Soft-LKG, and COMIX-style selection attain intermediate QoS-satisfaction ratios, whereas CTO-only and the naive blend satisfy only one of the two QoS targets on average.

\begin{table}[!t]
\caption{QoS-Satisfaction Ratio Across Operator Priorities}
\label{tab:qos}
\centering
\renewcommand{\arraystretch}{1.2}
\begin{tabular}{lc}
\hline
\textbf{Method} & \textbf{QoS Satisfaction}\\
\hline
ES-only & $1.000 \pm 0.000$\\
QACM-style & $1.000 \pm 0.000$\\
Soft-ES & $0.742 \pm 0.060$\\
Proposed hard-switch & $0.740 \pm 0.061$\\
Soft-LKG & $0.739 \pm 0.061$\\
COMIX-style & $0.724 \pm 0.066$\\
Naive blend & $0.500 \pm 0.000$\\
CTO-only & $0.500 \pm 0.000$\\
\hline
\end{tabular}
\end{table}

The proposed method is not designed to maximize QoS-threshold satisfaction. Instead, it optimizes the energy-aware utility of \eqref{eq:4}. Consequently, methods optimized for QoS targets, such as ES-only and QACM-style arbitration, achieve higher QoS-satisfaction ratios while incurring larger utility regret. This difference reflects distinct optimization objectives rather than a deficiency of either approach. Reporting both metrics therefore provides a more balanced assessment of conflict-resolution behavior.

\subsection{Sensitivity to the EWMA Factor}
\label{subsec:betasens}

Table~\ref{tab:betasens} summarizes the sensitivity of the proposed method to the EWMA factor $\beta$. Although $\beta=0.70$ achieves the lowest average regret, the differences among the tested values remain small.

\begin{table}[!t]
\caption{EWMA-Factor Sensitivity: Overall Hard-Switch Utility Regret for
$\beta \in \{0.5,0.7,0.9\}$ ($\tau=1$; Mean $\pm$ 95\% CI over Seeds
and Drift). Bold Marks the Reported Value $\beta=0.70$.}
\label{tab:betasens}
\centering
\renewcommand{\arraystretch}{1.2}
\begin{tabular}{cccc}
\hline
$w_{\mathrm E}$ & $\beta=0.5$ & $\beta=0.7$ & $\beta=0.9$\\
\hline
0.1 & $0.482 \pm 0.194$ & $\mathbf{0.466 \pm 0.192}$ & $0.548 \pm 0.221$\\
0.2 & $0.496 \pm 0.160$ & $\mathbf{0.448 \pm 0.137}$ & $0.475 \pm 0.131$\\
1.0 & $0.544 \pm 0.177$ & $\mathbf{0.543 \pm 0.176}$ & $0.579 \pm 0.188$\\
\hline
\end{tabular}
\end{table}

The observed variation is negligible compared with the robustness gains achieved over the COMIX-style baseline under significant drift. The slightly higher regret observed at $\beta=0.9$ is consistent with the slower drift-detection dynamics predicted by \eqref{eq:15}. Overall, the results indicate that the proposed method is not highly sensitive to the precise choice of $\beta$.

\subsection{Key Findings}

We identify three main observations from our evaluation. 
First, the utility-maximizing transmit-power compromise depends strongly on the operator energy weight. Consequently, fixed arbitration policies cannot provide uniformly optimal behavior across deployment objectives.
Second, NDTs are valuable decision-support tools but should not be treated as intrinsically reliable. Blind twin-based arbitration performs well when the twin is accurate but deteriorates rapidly under drift as it continues to trust biased predictions.
Third, explicit fidelity monitoring provides an effective safeguard against such degradation. Across interior-optimum regimes, the proposed method consistently maintains low regret by switching from twin-recommended actions to live-validated actions when prediction accuracy deteriorates. At 10~dB drift, this mechanism reduces regret by more than one order of magnitude relative to blind twin-based selection. In boundary-optimum regimes, where drift does not materially affect the optimal operating point, the proposed method matches the strongest baselines rather than introducing a performance penalty.
These observations indicate  that online NDT fidelity monitoring is a lightweight and effective mechanism for robust utility-aware arbitration of direct O-RAN xApp conflicts under model drift.

\section{Conclusion}
\label{sec:conclusion}

This paper investigated resolving direct xApp conflicts in O-RAN when arbitration decisions rely on an NDT whose accuracy may degrade over time. Focusing on a representative conflict between energy-saving (ES) and coverage/throughput-oriented (CTO) xApps, we formulated conflict resolution as a utility-aware transmit-power arbitration problem and proposed a twin-fidelity-aware arbitration mechanism. The proposed approach combines twin-based action evaluation with online fidelity monitoring and a lightweight fallback strategy based on a live-validated LKG action. By continuously tracking the mismatch between predicted and realized utility, the arbiter can exploit the twin when it remains reliable and automatically reduce its influence when drift is detected. The resulting mechanism is training-free, model-agnostic, and does not require modifying, retraining, or coordinating the underlying xApps.
Through system-level evaluation, we demonstrated three key findings. First, the utility-maximizing conflict-resolution decision depends strongly on the operator's energy-efficiency preference, confirming the need for adaptive arbitration instead of fixed policies. Second, blind reliance on NDT recommendations can lead to substantial performance degradation as twin drift increases. Third, explicit fidelity monitoring provides an effective safeguard against degradation, allowing the proposed arbiter to maintain low utility regret across a wide range of drift conditions while matching the performance of the strongest baselines in regimes where fidelity monitoring offers limited additional benefit. These results show that NDT fidelity is an important runtime signal that should be incorporated into O-RAN conflict-management decisions rather than assumed to remain constant throughout operation. Finally, this work highlights the importance of coupling AI-driven decision support with runtime trust assessment. While demonstrated on a direct transmit-power conflict, the underlying fidelity-aware arbitration principle is applicable to other O-RAN control parameters and to broader classes of xApp interactions. Future work will extend the framework to indirect and implicit conflicts, multi-parameter control actions, and non-stationary environments in which the optimal operating point evolves over time.

\balance
\bibliographystyle{IEEEtran}
\bibliography{refs}
\end{document}